\documentclass[format=acmsmall, review=false, screen=true]{acmart}

\usepackage{graphicx}  
\usepackage{float}  
\usepackage{bookmark}
\usepackage{pifont}  
\usepackage[T1]{fontenc}  
\usepackage[utf8]{inputenc}  
\usepackage{csquotes}
\usepackage{setspace}
\usepackage{textcomp}  
\usepackage[load=abbr]{siunitx}  
\usepackage{listings}
\usepackage{booktabs} 

\usepackage[ruled]{algorithm2e} 

\SetAlFnt{\small}
\SetAlCapFnt{\small}
\SetAlCapNameFnt{\small}
\SetAlCapHSkip{0pt}
\IncMargin{-\parindent}

\setcopyright{none}

\received{April 2019}
\received[revised]{-}
\received[accepted]{-}

\newcommand{\cmark}{\ding{51}}%
\begin{document}
\title[PeyeDF: an Eye-Tracking Application]{PeyeDF: an Eye-Tracking Application for Reading and Self-Indexing Research}

\author{Marco Filetti}
\orcid{0000-0002-2094-2201}
\affiliation{%
  \institution{University of Helsinki}
  \department{Helsinki Institute for Information Technology HIIT}
  \streetaddress{Gustaf Hällströmin katu 2b}
  \city{Helsinki}
  \state{Uusimaa}
  \postcode{00014}
  \country{Finland}}
\email{marcofiletti@outlook.com}

\author{Hamed R. Tavakoli}
\affiliation{%
  \institution{Aalto University}
  \department{Department of Computer Science}
  \city{Espoo}
  \country{Finland}
}
\email{hamed.r-tavakoli@aalto.fi}

\author{Niklas Ravaja}
\affiliation{%
  \institution{Aalto University}
  \department{Department of Information and Service Economy}
  \city{Helsinki}
  \country{Finland}
}
\affiliation{%
  \institution{University of Helsinki}
  \department{Department of Social Research}
  \city{Helsinki}
  \country{Finland}
}
\email{niklas.ravaja@aalto.fi}

\author{Giulio Jacucci}
\affiliation{%
  \institution{University of Helsinki}
  \department{Helsinki Institute for Information Technology HIIT}
  \city{Helsinki}
  \country{Finland}
}
\email{giulio.jacucci@helsinki.fi}

\begin{abstract}

PeyeDF is a Portable Document Format (PDF) reader with eye tracking support, available as free and open source software. It is especially useful to researchers investigating reading and learning phenomena, as it integrates PDF reading-related behavioural data with gaze-related data. It is suitable for short and long-term research and supports multiple eye tracking systems. We utilised it to conduct an experiment which demonstrated that features obtained from both gaze and reading data collected in the past can predict reading comprehension which takes place in the future. PeyeDF also provides an integrated means for data collection and indexing using the DiMe personal data storage system. It is designed to collect data in the background without interfering with the reading experience, behaving like a modern lightweight PDF reader. Moreover, it supports annotations, tagging and collaborative work. A modular design allows the application to be easily modified in order to support additional eye tracking protocols and run controlled experiments. We discuss the implementation of the software and report on the results of the experiment which we conducted with it.

\end{abstract}




\maketitle

\renewcommand{\shortauthors}{Filetti et al.}


\section{Introduction}\label{sec:introduction}

Eye tracking is a valuable, non-invasive means of investigation. It has seen increased adoption rates in the past decade, thanks to gradual but steady technological and cost-related improvements \cite{jacob2003eye}. Many fields of research have been taking advantage of its capabilities, ranging from design \cite{djamasbi2010}, collaboration \cite{duchowski2002}, learning \cite{gutl2005adele}, cognition \cite{Blair2009,lalle2016predicting,wilson2003real} and usability \cite{jacob2003eye}. 

To date, the experiment design, experiment procedure and utilization of the eye tracking devices have been a laborious task for users and researchers, in particular for a long run experiment. In many circumstances, most of the users fail having a tangible connection with the software user interface and the researchers have to spend huge amount of time on implementing the experiments using eye tracking software. In the context of providing an appealing user experience and easier experiment design procedure, in particular for long run experiments, we present PeyeDF, an open-source eye-tracking enabled reading tool.
PeyeDF is suitable for the conduction of eye-tracking-based reading and learning experiments conducted over long periods of time. For users, PeyeDF is easy to use, lightweight, stable software that provides support for storing easily-performed manual document annotations. For experimenters, it provides out-of-the box support for multiple eye tracking platforms. 

\subsection{Related work}

Deciphering the state of one's mind by performing inferences on eye movements is a research topic that has been undertaken for various purposes. The span of this research area is wide-range and includes observer task inference \cite{borji2014}, tagging image pleasantness \cite{Tavakoli2015}, inferring traits \cite{hoppe18}, predicting one's age \cite{Oli18}, information retrieval \cite{Tavakoli16}, etc. Reviewing all the related work is beyond the intent of this article, for brevity, we refer the readers to \cite{duchowski2002}. A brief account of the most relevant work is, however, presented.

Investigating the relationship between eye movements and learning can provide great insight into cognitive functions.
To investigate one's cognitive performance during learning, reading has been a central focus in conjunction with the eye tracking. For example, it has been demonstrated that the duration and number of fixations increase along with the difficulty of the text being read \cite{rayner2006comprehension}. Such fixation-level measurements is utilised to assess how sentence construction affects comprehension difficulty \cite{traxler2002clauses} or to estimate the reader's language expertise \cite{kunze2013towards}. More examples are given in a review by \cite{gonzalezmarquez2007eye}. 

Within the concept of learning via “Lifelogging”, storing our data in personal databases \cite{gurrin2014} and retrieving previously manually annotated information is demonstrated to reinforce memory \cite{o1998student}. Annotations might also be performed automatically with attentive documents \cite{Buscher2012}. Both are valid strategies for the location of important passages that exploit manually or implicitly (automatically) annotated information. One drawback of automated annotation approaches is that the reinforcement of learning induced by the act of manually annotating passages is lost. In this context, the possibility of manually annotating passages is a feature that must be provided by all learning platforms \cite{adler1998diary}. The PeyeDF tries to alleviate this shortcoming by allowing both manual and implicit annotation schemes.

Electronic devices have introduced new research questions for learning, such as the investigation of screen-based reading behaviour. There are opposing views in favour \cite{chen2014,Liu2005} and against such technologies \cite{ackerman2011,mangen2013} in learning and education. Many factors such as meta-cognitive abilities, culture, familiarity, habit, improved display technologies and age are argued to influence performance of users in digital reading \cite{taipale2014,franze2014,ball2011rethinking}. While electronic reading is gradually replacing paper-based reading \cite{adler1998diary,Liu2005}, the reading experience is key to success in reading. In other words, being comfortable with digital tools can improve learning \cite{harrison2000ebooks}. We, thus, need to have well-designed software, akin to PeyeDF, for learning studies.

Designing software for the collection of user data during reading and learning can be cumbersome; a number of software systems have been developed to facilitate such tasks. These are summarised in \autoref{tab:relatedsoftware}. 

Adele is a e-learning suite that allows synchronisation of eye tracking data with course material \cite{gutl2005adele}. Text 2.0 \cite{biedert2010text} and Eyebook \cite{Biedert2010} are two frameworks for web browsers designed to provide interactive feedback to gaze data. These two systems, along with that presented by \cite{campbell2001robust}, can distinguish between skimmed versus read text. Eye an Pen \cite{Alamargot:2006aa} enables synchronisation of handwriting with gaze data. Ogama is a system for slideshow study designs \cite{vosskuhler2008}. iTrace is an environment that enables eye tracking research on software systems, rather than static text \cite{shaffer2015itrace}. Eye tracking software prototypes have also been developed for tablets and smartphones, although these are limited in accuracy and speed due to current hardware \cite{wood2014eyetab,kunze2013my}. 

Task Tracer is a Windows application for tracking user activity across various tasks \cite{dragunov2005tasktracer}. It provides access to an open database that stores computer usage data. “Stuff I’ve Seen” is a system for personal data storage and retrieval \cite{dumais2016stuff}. Presto is a document storage system that organises material based on its contents, in addition to the traditional folder-based approach \cite{dourish1999presto}. Xlibris is a tablet-based proprietary system for document storage and annotation \cite{Schilit1998}. Note that some these applications do not support eye tracking and are not open-source. They are, thus, not among the best choices for researchers.

Among the open-source applications, Varifocal Reader is an application that allows navigation of large text documents at various level of detail (e.g. headers, sub-headers) \cite{Koch2014}. It is aimed at exploring individual large documents (books), while PeyeDF is aimed at storing information about collections of documents of any size. GazeParser is an open-source library for low-cost eye tracking and data analysis; it consists of a video-based eyetracker and libraries \cite{sogo2013agazeparser}. It is not aimed at reading nor does it support long-term data storage and retrieval. Eyerec \cite{santini2016eyerec} and openEyes \cite{li2006openeyes} are two open-source hardware specification for eye tracking.

To summarize, the current eye tracking-based systems, in one hand, do not support personal data storage (PDS) and can not support lifelogging. On the other hand, the PDS based systems do not support any eye tracking and can not be used in eye tracking research. The PeyeDF software addresses these shortcommings by providing support for eye tracking and PDS. Moreover, it is an open source, lightweight PDF reader that can be customized for 
numerous research purposes.


\begin{table}[htb]
    \caption{Comparison of software related to PeyeDF}
    \label{tab:relatedsoftware}
    \tabcolsep=0.15cm
    \begin{tabular}{lllccc}
        \hline
        Name & Reference & Purpose & Eye Tracking & PDS* & Open-source \\
        \hline
        Adele & \cite{gutl2005adele} & Learning & \cmark &  &  \\
        Text 2.0 & \cite{biedert2010text} & Reading & \cmark &  &  \\
        Eyebook & \cite{Biedert2010} & Reading & \cmark &  &  \\
        Eye and Pen & \cite{Alamargot:2006aa} & Learning & \cmark &  &  \\
        Ogama & \cite{vosskuhler2008} & Presentations & \cmark &  &  \\
        iTrace & \cite{shaffer2015itrace} & Programming & \cmark &  &  \\
        Eyetab & \cite{wood2014eyetab} & Tablets & \cmark &  &  \\
        My Reading Life & \cite{kunze2013my} & Phones & \cmark &  &  \\
        Varifocal Reader & \cite{Koch2014} & Large documents & \cmark &  & \cmark \\
        Task Tracer & \cite{dragunov2005tasktracer} & Office work &  & \cmark &  \\
        Stuff I've Seen & \cite{dumais2016stuff} & PDS &  & \cmark &  \\
        Xlibris & \cite{Schilit1998} & Tablets &  & \cmark &  \\
        GazeParser & \cite{sogo2013agazeparser} & Software library & \cmark &  & \cmark \\
        Eyerec & \cite{santini2016eyerec} & Hardware & \cmark &  & \cmark \\
        Openeyes & \cite{li2006openeyes} & Hardware & \cmark &  & \cmark \\
        \hline
        PeyeDF & ~ & Reading & \cmark & \cmark & \cmark \\
        \hline
    \end{tabular}
    {*Supports or includes a Personal Data Storage system.}
\end{table}

\subsection{Contribution}

We introduce PeyeDF. It is an open-source eye-tracking enabled reading tool. The PeyeDF is designed to support personal data storage systems (PDS) and eye tracking support in
combination with a lightweight document viewer that enhances user experience during learning experiments. Some of the unique properties of the PeyeDF includes,

\begin{itemize}
\item tracking viewport position,
\item tracking zoom factors,
\item support for multiple eye tracking devices,
\item implicit data annotation
\item extendable modules,
\item supporting DiMe, a personal data storage system,
\item lightweight and user friendly interface
\end{itemize}

The PeyeDF eases the life of both experimenters and users in long run reading and learning experiments. It has been tested over a long (3+ years) at Aalto University, Finland and to run the experiment described in \autoref{sec:experiment}. It can be also used as a stand alone self-indexing system when an eye tracker is not available.

\section{Implementation}\label{sec:implementation}

PeyeDF is available only for macOS and it has been highly optimised for the platform. All computation related to eye tracking data and behavioural data collection takes place in the background, so that the main thread (which updates the User Interface) is left unburdened, resulting in a smooth user experience; this is enabled by the implementation described in \autoref{sec:implementation}. It also supports spotlight (described in \autoref{sec:spotlightintegration}) and URL Types (explained in \autoref{sec:urltypes}). It implements the Swift API guidelines\footnote{\url{https://swift.org/documentation/api-design-guidelines/}} and the Apple Human Interface guidelines\footnote{\url{https://developer.apple.com/design/human-interface-guidelines/macos/overview/themes/}}.

\subsection{Installation}

\begin{figure}
\centering
\includegraphics[width=\columnwidth]{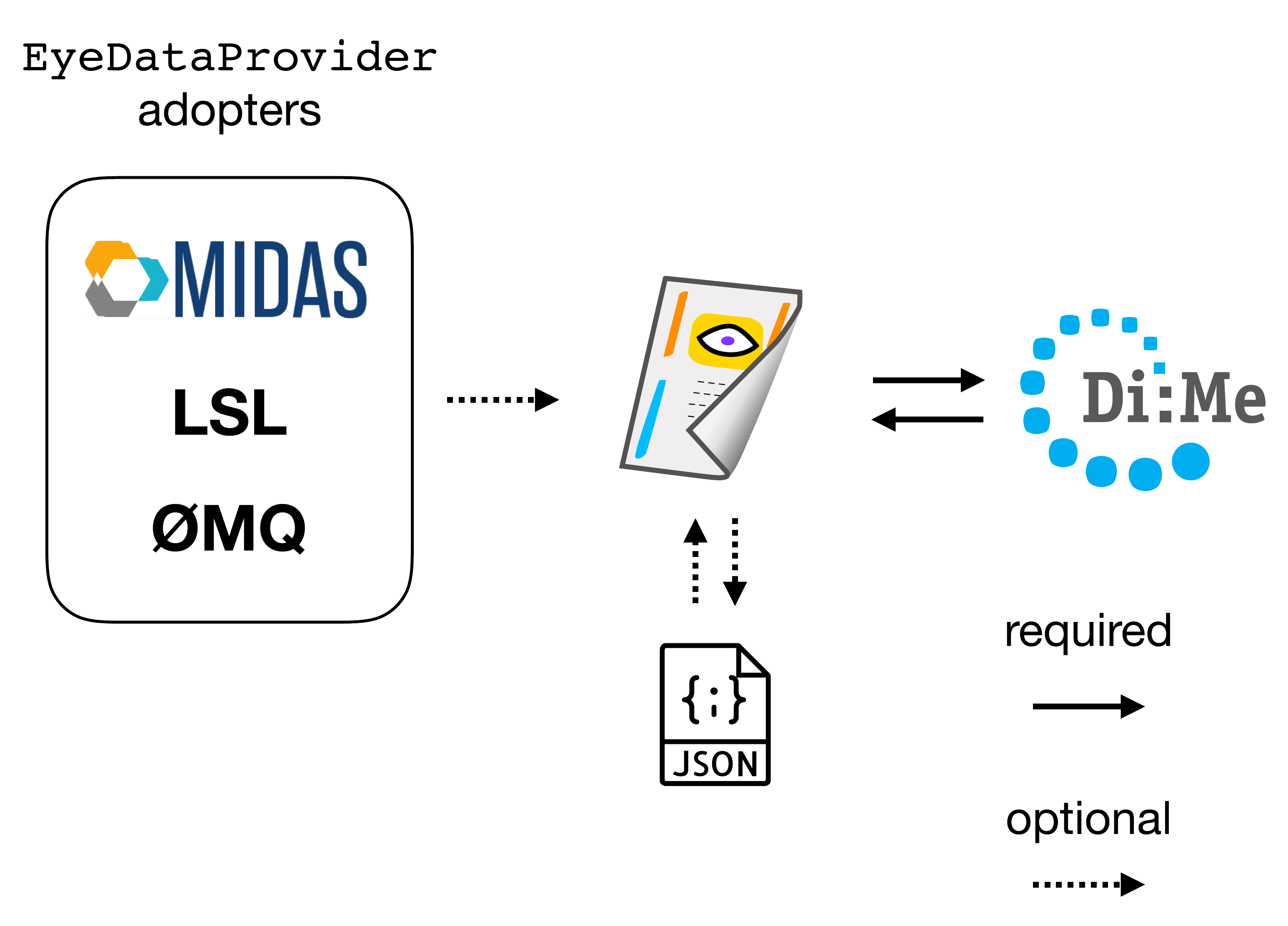}
\caption{System overview. PeyeDF requires two way communication with DiMe. It supports the listed protocols for eye tracking communication. As discussed in \autoref{sec:generalisability}, the \texttt{EyeDataProvider} protocol generalises eye tracking support. It is also possible to export / import eye tracking and reading data via JSON files. } \label{fig:overview}
\end{figure}

PeyeDF is a standard macOS application; as such, it can be installed by downloading a disk image from \url{https://github.com/HIIT/PeyeDF/releases} and dragging PeyeDF to the Applications folder. However, in order to function it requires the installation of DiMe and, optionally, a connection to a supported eye tracker. These steps are described below. An overview of the full system is given in \autoref{fig:overview}.

\subsection{Prerequisites}

PeyeDF requires DiMe for data storage and retrieval. DiMe is an open-source Personal Data Storage (PDS) system that supports multiple applications and data types \cite{symbiotic2016}. DiMe itself requires the following software to be installed in order to be compiled.

\begin{itemize}
    \item Xcode\footnote{\url{https://itunes.apple.com/app/xcode/id497799835}} or Command Line Tools \footnote{\texttt{xcode-select {-}{-}install}}
    \item Java SE\footnote{\url{http://www.oracle.com/technetwork/java/javase/downloads}} JDK version 8 or above
    \item Node.js, downloadable from the web\footnote{\url{https://nodejs.org}} or via Homebrew\footnote{\texttt{brew install node}}, if installed
\end{itemize}

\subsection{Setup}

Once the required software is installed, DiMe should be compiled and run. PeyeDF must then be configured to utilise DiMe using a predefined user and password. The procedure to do so is detailed below and the related code can be pasted into a Terminal (note that individual commands are separated by~‘\texttt{;}’ and otherwise each statement should be written in a single line).

\begin{enumerate}
    \item Clone the open-source DiMe repository: \\
\texttt{git clone {-}{-}recursive https://github.com/HIIT/dime-server}
    \item Run DiMe: \\
\texttt{cd dime-server; make run}
    \item DiMe is ready Once ‘fi.hiit.dime.Application: Started’ appears; first-time compilation may take a few minutes.
    \item Navigate to \url{http://localhost:8080} and create a user. For a test run, use Test1 as user and 123456 for password as these are the PeyeDF defaults.
    \item PeyeDF can now be started and should be fully functional. If a different user rather than the suggested Test1 was created during the previous step, navigate to PeyeDF Preferences, select the ‘DiMe’ tab and enter the chosen username and password.
\end{enumerate}

Dime creates a \texttt{.dime} directory under the user's home. This directory contains the full database; deleting this directory will reset DiMe. If the directory is copied or moved to another machine an installation of DiMe on that machine will be able to read the database (assuming that username and password match). DiMe can also be installed on a (local) network server so that it can be used by multiple users simultaneously.

\section{Usage}\label{sec:usage}

\begin{figure*}
\centering
\includegraphics[width=\textwidth]{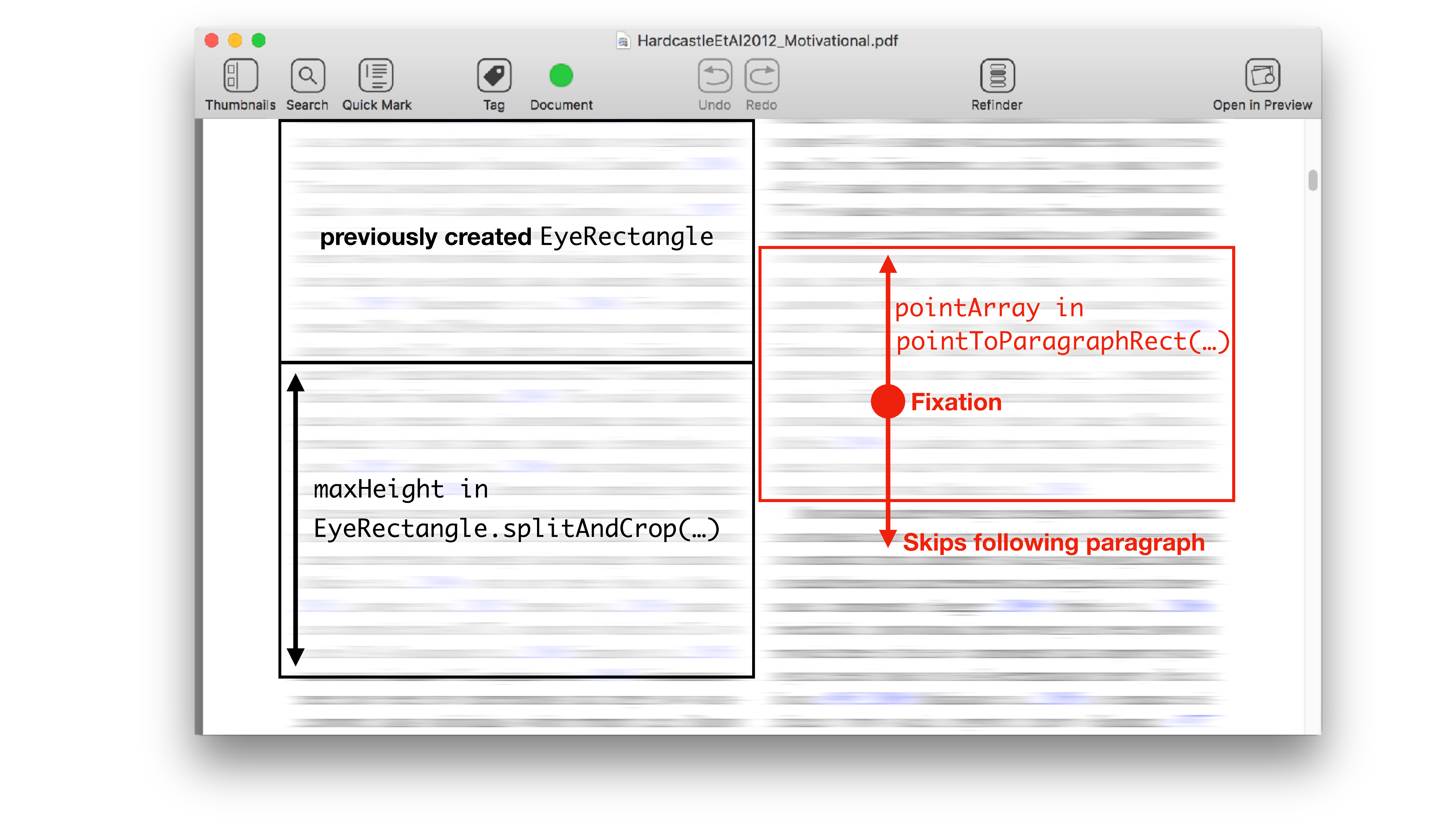}
\caption{Screenshot and paragraph detection. This figure shows a screenshot of PeyeDF (an instance of \texttt{DocumentWindowController}) with overlays representing fixations and their related paragraphs. A fixation is assigned to a paragraph when its height can be contained within 3$\degree$ of visual angle from the fixation. PeyeDF stores all fixation and paragraph information in sets of \texttt{ReadingEvent}s, as described in \autoref{sec:events}.} \label{fig:screenandevent}
\end{figure*}

PeyeDF works and behaves like a lightweight PDF reader -- similarly to the stock ‘Preview’ application on macOS. PDF files can be opened with PeyeDF by default, if desired. Eye tracking support is not required for the application to work, although it is the main motivation for using PeyeDF, along with its data storage and indexing capabilities. A screenshot of the applications is provided in \autoref{fig:screenandevent}, which also shows how PeyeDF captures the currently read paragraph while using an eye tracker.

PeyeDF supports two eye tracking hardware systems: SMI\footnote{\url{https://www.smivision.com}} and Pupil Labs\footnote{\url{https://www.pupil-labs.com}}. Hardware support is enabled via data input from the LSL\footnote{\url{https://github.com/sccn/labstreaminglayer}}, zeroMQ\footnote{\url{http://zeromq.org}} and MIDAS\footnote{\url{https://github.com/bwrc/midas}} protocols. If desired, PeyeDF can be extended to support additional protocols (and hence hardware) as discussed in \autoref{sec:generalisability}. To activate eye tracking while using PeyeDF, it is sufficient to select the desired protocol in the Preferences > Experiment dialog. An error message will be displayed if a valid connection was not be established. All gaze data is fetched in real time; at this stage, PeyeDF does not support off-line gaze recorders such as GazeParser \cite{sogo2013}. If eye tracking is currently being used, a tick for ‘Eye Tracking’ appears in the ‘Connections’ menu.

\begin{figure}
\centering
\includegraphics[width=\columnwidth]{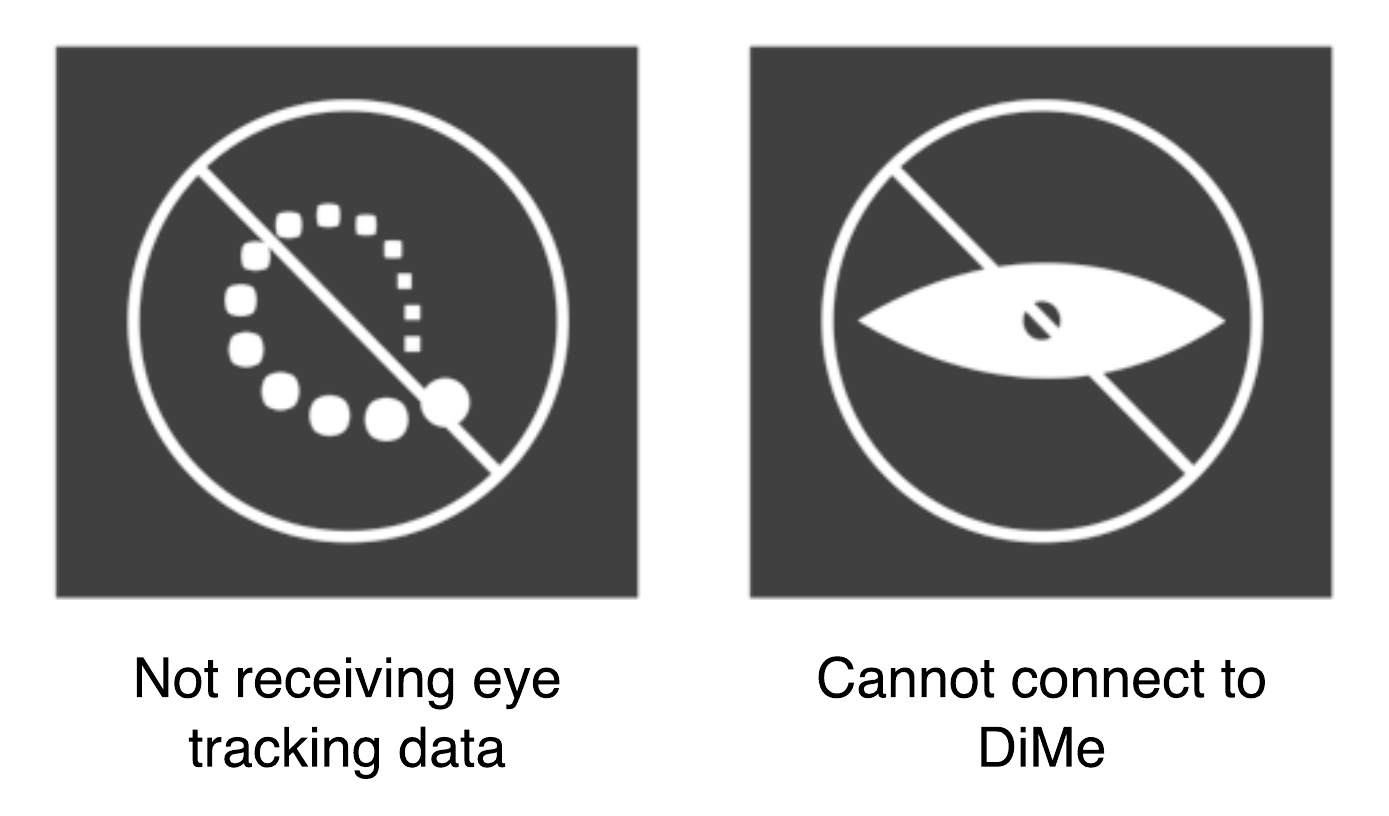}
\caption{Error indicators. These images are shown over the reading area when a connection error is detected, in order to capture the user's attention. The `Not receiving eye tracking data' image is also shown when the user's eyes have not been detected for an extended period of time (e.g. in case the user left the workstation).} \label{fig:errors}
\end{figure}

In the ‘Connections’ menu, a tick also indicates whether PeyeDF is successfully connected to DiMe. In case a connectivity problem occurs during usage, one of the images displayed in \autoref{fig:errors} is displayed over the reading area to attract the user's attention.

PeyeDF collects eye tracking and reading data in the background, without noticeably affecting the reading experience. This is enabled by asynchronous processing of behavioural and eye tracking data in background queues (as described in \autoref{sec:performance}).

PeyeDF provides support for rapid annotations. Annotation support is critical for reading applications as actively highlighting passages can increase their retention \cite{Fowler1974}. It has been previously suggested that annotations in the digital world may be less effective than paper-based annotation when they are too complex to use \cite{benyehudah2014influence}. Providing a means for quickly annotating sections of relevant text is then beneficial. To enable quick annotations, the `quick mark' button (visible on the toolbar) should be activated. Annotations are then performed with a double-click on a paragraph to mark an “important” paragraph in yellow. A triple click would highlight the area as “critical”, using a stronger colour (red). It is also possible to annotate text by selecting a range of text and right clicking.

Tagging is also supported: users can assign tags (corresponding to short strings) to any range of text. Tags are stored in DiMe, and can subsequently searched and retrieved by PeyeDF or other supporting applications.

\subsection{Refinder}

It has been suggested that lack of visual cues in digital applications might impede reading \cite{ackerman2011,mangen2013}. PeyeDF implements a Refinder functionality, which utilises colours, tags and overviews in order to provide spatiotemporal markers to previously read and annotated text. The Refinder has been included in PeyeDF for two reasons. Firstly, it improves the reading experience by providing a simple means to revisit previously highlighted text; this may motivate experiment participants in utilising PeyeDF rather than a default reader. Secondly, the Refinder has been designed in order to be extensible: it can be integrated into third party applications (as described in \autoref{sec:urltypes}) or it can be modified to investigate novel refinding approaches (all Refinder functionality is collected in the ‘Refinder’ folder in our repository\footnote{\url{http://github.com/HIIT/PeyeDF/tree/master/PeyeDF/Refinder}}).

\begin{figure*}
\centering
\includegraphics[width=\textwidth]{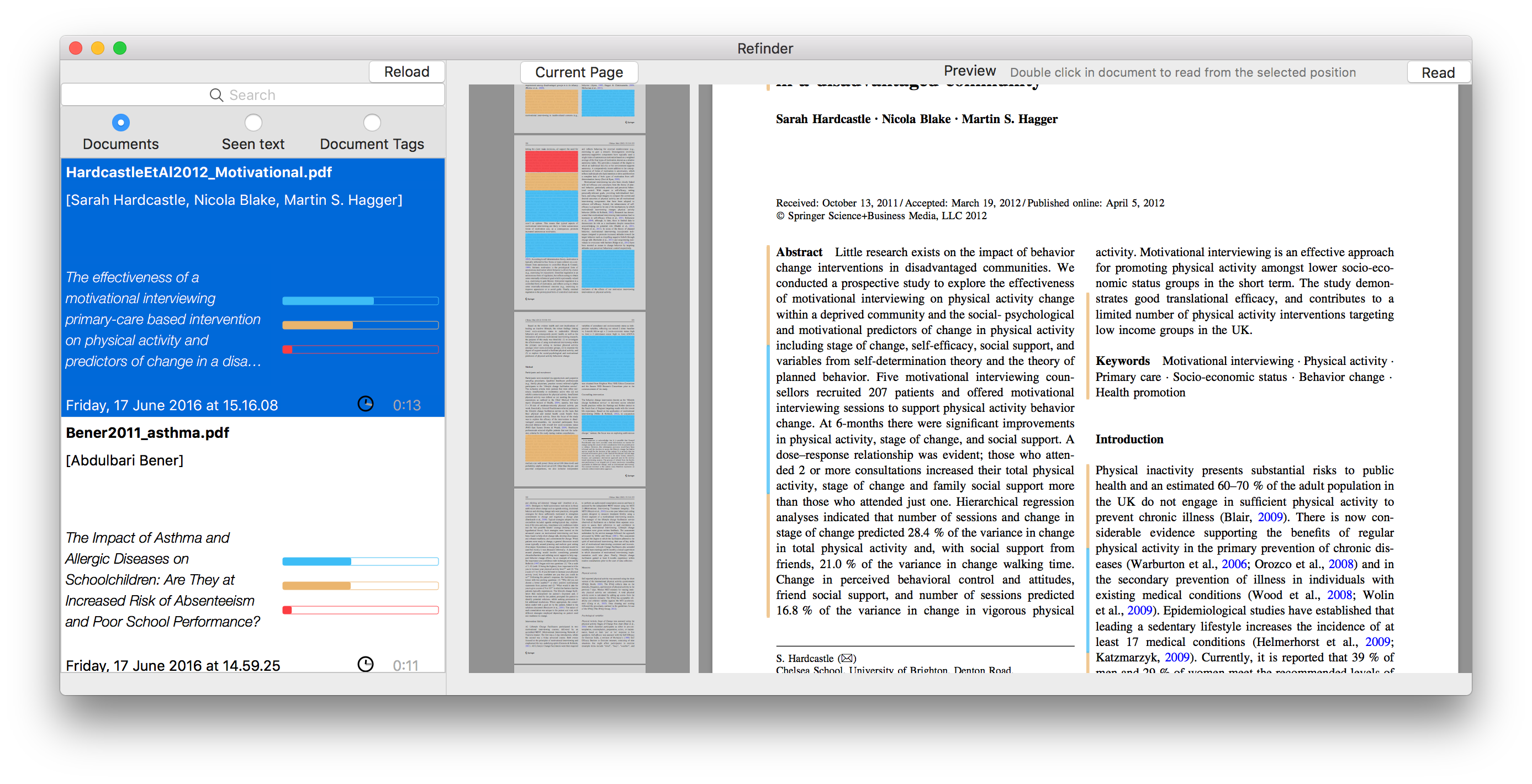}
\caption{Refinder screenshot. A screenshot of an instance of \texttt{RefinderWindowController} is shown in this figure. There can only be one such window in an instance of PeyeDF. Colours represent the type of annotations created by users (yellow for ``important'' and red for ``critical''). In this example, blue represents unannotated paragraphs which contained at least three fixations. The colour bars represent the overall proportion of the given document which is marked in the given colour. Note that in the overview, blocks are overlaid over text, while in the main window annotations are displayed as a bar on the left side of the related paragraph.} \label{fig:refinderscreenshot}
\end{figure*}

\subsection{Collaboration}

PeyeDF also supports collaborative reading and tagging. To enable it, two (or more) collaborators must be connected to the same local network. Detected collaborators can be displayed by activating the  Connections > Show network readers menu. One can then initiate collaboration by pressing the ‘invite’ button. Collaboration also requires DiMe. This allows users to ``track'' one another, if desired, showing exactly which paragraph is being read by the collaborator using eye tracking.

Real time sharing of gaze information has been shown to increase collaboration quality during problem solving \cite{schneider2013,stein2004another}, collaborative visual search \cite{brennan2008} and a planning task \cite{cherubini2010indicating}. It has also been shown to be useful, for example, in remote non-collaborative scenarios, such as virtual lectures \cite{sharma2015displaying} or virtual conferencing \cite{vertegaal1999gaze}. Support for collaborative work in reading software is a desirable feature according to \cite{adler1998diary}.

\begin{figure*}
\centering
\includegraphics[width=\textwidth]{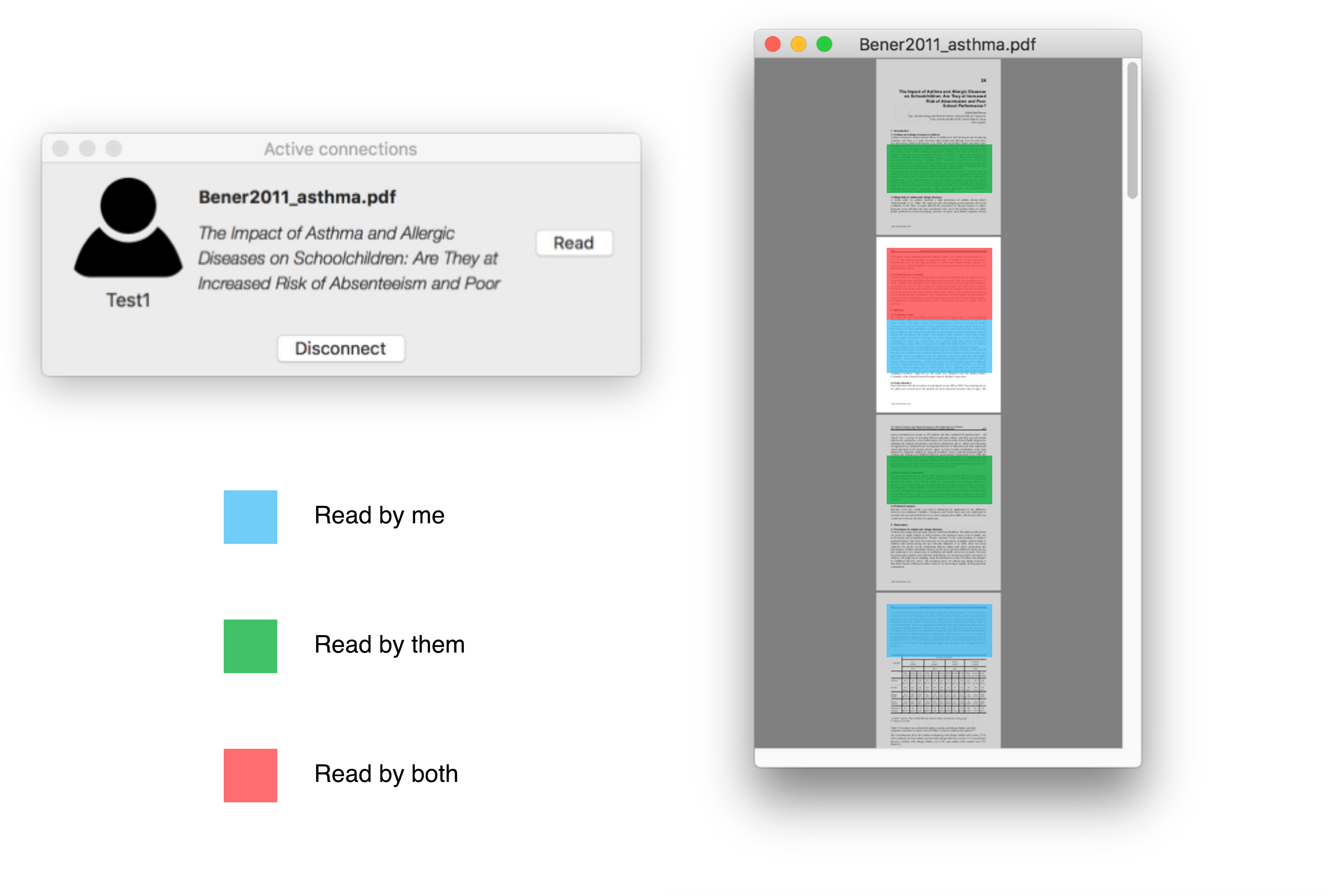}
\caption{Collaboration features. On the left, the list of active connections is shown; in this case, there is only one active connection. This displays the filename and title of the document currently being read by the collaborator(s). This item will also display a `Track' checkbox upon the start of a collaborative reading session; this allows one collaborator to follow the gaze and / or reading viewport of the other. On the right, the overview of the collaborative session. Colours identify sections of the document which have been read by the local user, the collaborator, or both. This overview is a small window displayed along with the main \texttt{DocumentWindowController}} \label{fig:collaboration}
\end{figure*}

Application elements unique to the collaboration feature are shown in \autoref{fig:collaboration}. This feature is provided to keep the application on par with modern office packages and potentially allows for research in collaborative environments, with and without eye tracking.

\subsection{Spotlight integration}\label{sec:spotlightintegration}

A minor but useful feature is spotlight integration. When searching for text on a macOS system, opening a PDF with PeyeDF will open it while searching for the previously searched text. Searched-for text is stored in \texttt{SummaryReadingEvents}. Events are explained in the next section.

\section{Data Structures and APIs}\label{sec:data}

\begin{figure*}
\centering
\includegraphics[width=\textwidth]{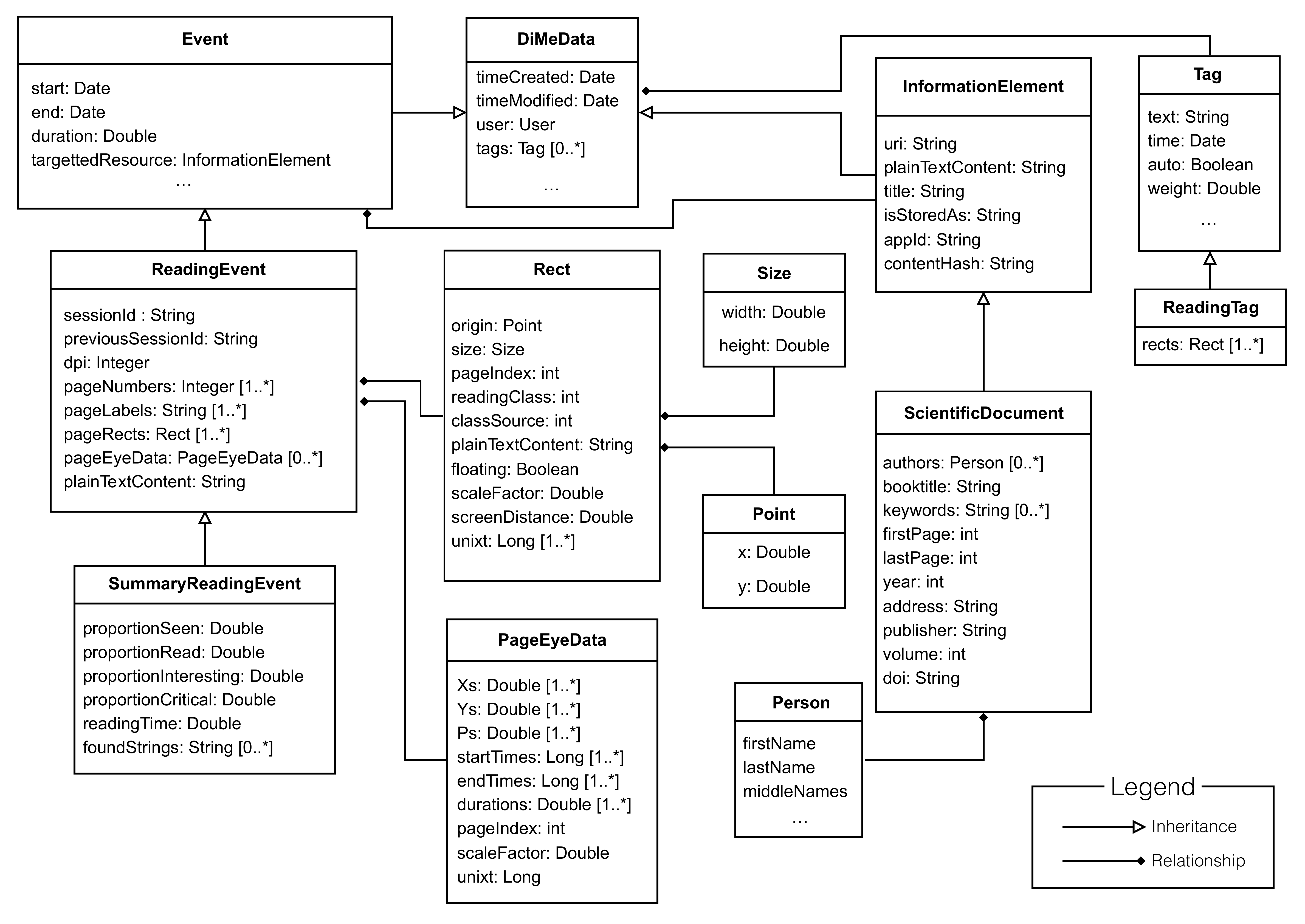}
\caption{Data class diagram. Shows the relationships and format of data classes that PeyeDF exports. DiMeData is top-level class from which all other inherit; this is used in PeyeDF to identify classes which can be stored in DiMe. Field names were designed to be self-explanatory. \texttt{Event} is the most relevant class for retrieving eye data (along with its subclasses); this is explained in \autoref{sec:events}. If needed, more details are provided, in the dime-server repository data folder$^{\ref{ft:dimedatafolder}}$ or the PeyeDF repository wiki$^{\ref{ft:dataformatwiki}}$.} \label{fig:dataclassdiagram}
\end{figure*}

This section describes the format of PeyeDF data structures, APIs, events and their usage. \autoref{fig:dataclassdiagram}
summarises the data types supported by DiMe. This information is also documented in the dime-server repository\footnote{\url{https://github.com/HIIT/dime-server/tree/master/src/main/java/fi/hiit/dime/data}\label{ft:dimedatafolder}} and
in the corresponding wiki pages\footnote{\url{https://github.com/HIIT/PeyeDF/wiki/Data-Format}\label{ft:dataformatwiki}}.

For a generic user, accessing the stored data of an experiment is likely the most critical functionality of the software. Data can be retrieved in two ways: (1) by right clicking an entry (session) in the Refinder and choosing ‘Extract JSON’, which allows to save all \texttt{ReadingEvents} related to the selected session in a JSON file, and (2) by utilising the DiMe API, to be described in \autoref{sec:dimeapi}. While the first approach is relatively easy, the second requires better understandings of the internals of the software, and it also enables one to further extend the software.

\subsection{Data Types: Events and Information Elements}\label{sec:events}

There exists two main data types, which should be distinguished from each other, (1) \texttt{Event}s and (2) \texttt{InformationElement}s. \texttt{Event}s correspond to snapshots of user activity, while \texttt{InformationElement}s represent documents or other information retrievable with a URI. In PeyeDF, there is only one value for \texttt{InformationElement}, which is \texttt{ScientificDocument}, a link to the PDF files to be retrieved. An \texttt{InformationElement} contains also a \texttt{ContentHash} and an \texttt{appId}. The first, a SHA hash of the textual content of a PDF, is used to identify textual similarity between PDFs. \texttt{appId} is simply a \texttt{ContentHash} with ‘PeyeDF\_’ prefixed to distinguish between PeyeDF and other applications. 

The data from experiments will be mostly stored in the \texttt{Event} and its variants. We, thus, dedicate most of this section to the \texttt{Event}s data type. The software consists of several event data structures, namely  \texttt{Event}, \texttt{ReadingEvent}, and  \texttt{SummaryReadingEvent}. The first and most important is \texttt{Event} that contains \texttt{Event.targettedResource} field which contains  the associated element of a given event, i.e., a pointer to \texttt{InformationElement} data type corresponding to the current event. 

During the software operation, when a user wants to access a document for reading, the reading event is triggered to handle the retrieval of the relevant document and starts recording the action of reading.  \texttt{ReadingEvent} contains data collected from the moment the user started the act of reading until the reading stopped. A \texttt{ReadingEvent} may last a few seconds or minutes. We assume reading action is stopped as soon as the window is scrolled or moved. A more detailed explanation of how we estimate whether the user is reading is provided in \autoref{sec:performance}. 
  
  Each \texttt{ReadingEvent} has a \texttt{sessionId} in terms of a string UUID in PeyeDF. During a reading session, the same \texttt{sessionId} is allocated to all \texttt{ReadingEvent}s of that specific reading session. A session starts when the user opens a PDF document and terminates when the PDF is closed. Normally, one session comprises many events. 

\begin{figure*}
\centering
\includegraphics[width=\textwidth]{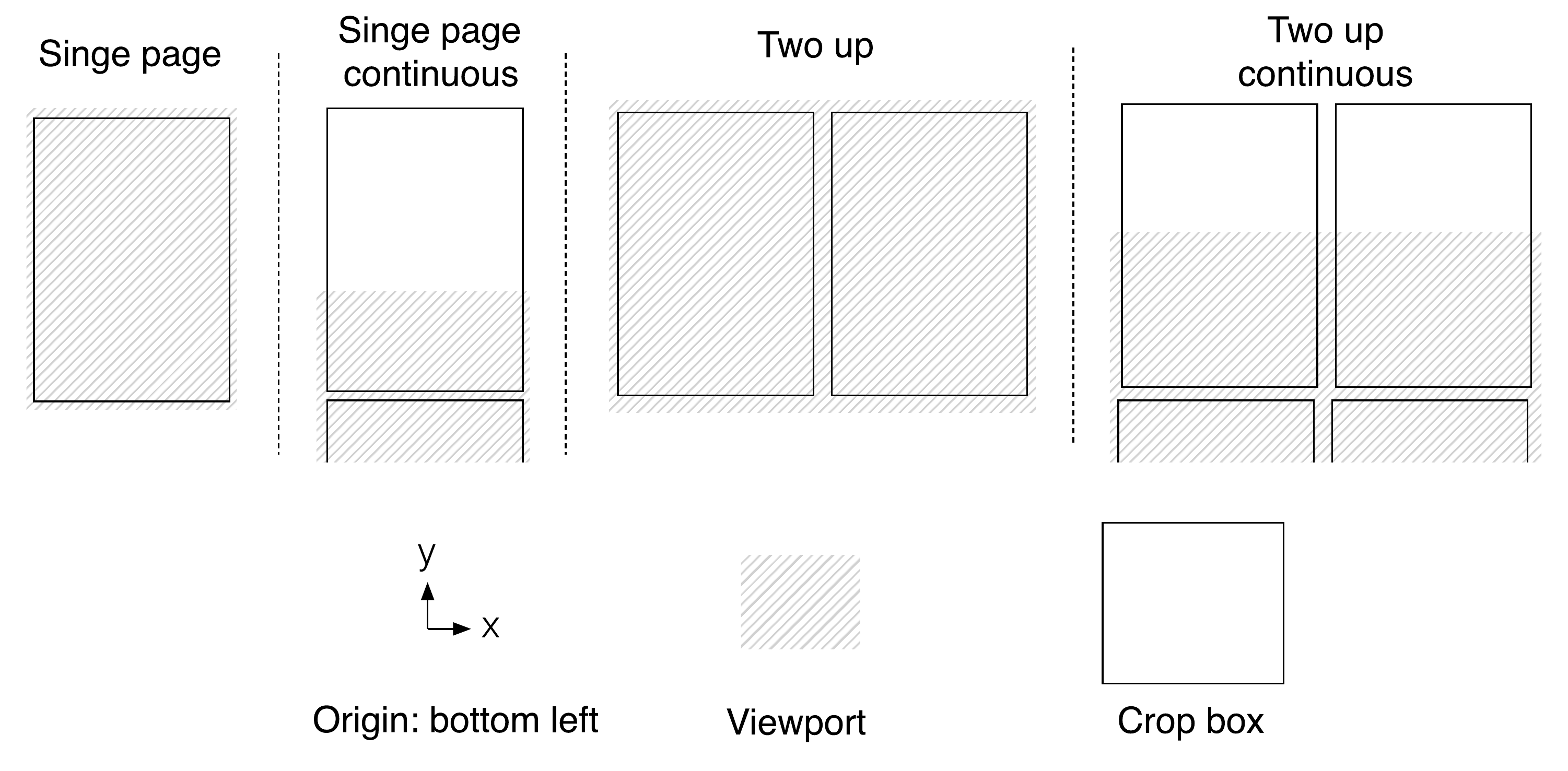}
\caption{Viewports. This figure displays the position of the viewport under different page arrangements. Note that origin is always at bottom left. When the viewport covers more than one page, the \texttt{pageLabels} and \texttt{pageNumbers} lists of a \texttt{ReadingEvent} will contain more than one value. The crop box represents the visible section of a PDF page (if a page has not been previously cropped, the crop box will correspond to the full page size). Viewports are specified using \texttt{Rect}s within each \texttt{ReadingEvent}. \texttt{Rect}s representing viewports are identified using a \texttt{readingClass} of 10 and a \texttt{classSource} of 1.} \label{fig:viewports}
\end{figure*}

A \texttt{ReadingEvent} has the following fields: \texttt{pageNumbers}, \texttt{pageLabels}, \texttt{pageRects} and \texttt{plainTextContent} to convey the visible information within a session. \texttt{pageNumbers} and \texttt{pageLabels} contains the same number of elements and represent which page(s) that were visible to user. Page number is intended to be a page index starting from 0, while page label is the page number written on the page extracted from the PDF. If page label can not be extracted from PDF source the same indexing as page index will be used. \texttt{plainTextContent} contains all text that was visible on in the viewport(s). \texttt{pageRects} is an array of \texttt{Rect}s, which represent positions of paragraph of various nature. These specify the viewport position when the event was generated and other paragraph-level information. Viewports are illustrated in \autoref{fig:viewports}. 

To distinguish \texttt{Rect}s of different nature, the two fields of \texttt{readingClass} and \texttt{classSource} are introduced. 
The \texttt{classSource} defines what is the source of a \texttt{Rect}, e.g.,
eye tracker, a click, a viewport, etc. \texttt{readingClass} contains the information about the importance of the event, e.g., an important pargraph vs. a regular one.

During eye tracking, a \texttt{ReadingEvent} can include one or several \texttt{PageEyeData} fields to contain information for fixation information.
The fixations' data is specified within multiple in-order arrays (e.g. horizontal and vertical positions, pupil size, duration). Their coordinates are specified in page space, with \texttt{pageIndex} identifying the page on which these took place. There is a \texttt{pageEyeData} entry for each page for which gaze data were detected. When using eye tracking, PeyeDF detects the paragraph of interest for the given fixation, using the implementation described in \autoref{sec:paragraphdetection}.

A \texttt{SummaryReadingEvent} is triggered once a document is closed, i.e., a reading session is terminated. In consequence, the PeyeDF computes a summary for the reading session and adds it to the database. This information will be further used for future retrievals. The summary information consists of all the searched keywords and search hits in a document, along with their statistical information that may be useful for further relevance information retrieval within documents.

\subsection{DiMe API}\label{sec:dimeapi}

DiMe is the central backbone of PeyeDF data storage. PeyeDF populates DiMe using HTTP Rest API calls (POST or DELETE) in JSON; these data can be subsequently retrieved by consumers using the GET calls described in this subsection. PeyeDF itself uses GET calls for displaying data within the Refiner. The full DiMe API and configuration instructions are described by \cite{symbiotic2016}.

Using the DiMe API, one can query the data structures discussed above.
The queries useful in the PeyeDF context are: \texttt{data/event}, \texttt{data/events}, \texttt{data/informationelement}, \texttt{data/informationelements}, \texttt{eventsearch} and \texttt{search}. The singular endpoints (\texttt{data/event} and \texttt{data/informationelement}) are used to retrieve a single event or element as JSON. These terminate with the unique integer id that one wants to retrieve (e.g. \texttt{data/informationelement/1} will fetch the first element stored in DiMe). The plural data endpoints retrieve 0 or more JSON objects and can be filtered using GET parameters. These are discussed more in detail in the next two subsections.

\subsubsection{data/events}

This endpoint is used to fetch \texttt{ReadingEvent}s. This is used when fetching reading data created by PeyeDF in the format described in \autoref{sec:data}. Parameters can be used to filter by \texttt{type}, \texttt{actor}, \texttt{sessionId}, \texttt{elemId} or \texttt{contentHash}. Query parameters can be chained.

Possible queries:

\begin{itemize}
    \item \texttt{type} allows to filter by type. PeyeDF uses two types: \texttt{\#ReadingEvent} and \texttt{\#SummaryReadingEvent}. Note that the type has to be prefixed by \path{http://www.hiit.fi/ontologies/dime/} (this is to distinguish between types created by different organisations).
    \item \texttt{actor} filters by the “creator” of the event. In the case of PeyeDF, there is only one actor: PeyeDF.
    \item \texttt{sessionId} as previously indicated, each reading session, per document (from window open to window close) is assigned one sessionId. This query parameter can be used to fetch all events generated during a single session. This is an alphanumeric string.
    \item \texttt{elemId} filters only events related to a given information element Id (an integer). This is used to obtain all events related to an individual PDF file.
    \item \texttt{contentHash} retrieves all events which were related to a specific file. Using hashes to identify files allows them to be found after renaming or moving.
\end{itemize}

\paragraph{Example}

A full query could then be \path{http://localhost:8080/api/data/events?type=http://www.hiit.fi/ontologies/dime/#ReadingEvent&actor=PeyeDF}

This query will fetch all events created by PeyeDF. The \texttt{targettedResource} field within each returned event is used to identify the document related to a given event. This allows one to parse the list while searching for specific metrics and relate the given event(s) to a document.

\subsubsection{data/informationelements}

This endpoint fetches Scientific Documents (which are the only type of information element used by PeyeDF). This endpoint is used to fetch document(s) with a specific \texttt{contentHash}, of a specific \texttt{type} or with a given \texttt{tag}.

\begin{itemize}
    \item \texttt{contentHash} is an alphanumeric string uniquely identifying the document or file.
    \item \texttt{type}; PeyeDF uses only type: \texttt{\#ScientificDocument}. As with event queries, PeyeDF prefixes \texttt{type} with \path{http://www.hiit.fi/ontologies/dime/}.
    \item \texttt{tag} allows to retrieve only documents for which a  user assigned the given tag. The tag can appear in any part of the document.
\end{itemize}

\paragraph{Example}

\path{http://localhost:8080/api/data/informationelements?type=http://www.hiit.fi/Fontologies/dime/#ScientificDocument&tag=done} fetches all \texttt{ScientificDocument}s (PDF files) that the user tagged with `done'.

\subsubsection{eventsearch}

To perform textual searches across all events, use the \texttt{eventsearch} endpoint. The query requires a parameter for the given query; optionally, any of the event filter parameters described above can be used.
 
 \paragraph{Example}
  
 \path{http://localhost:8080/api/eventsearch?query=pipe&type=http://www.hiit.fi/ontologies/dime/#ReadingEvent} looks for \texttt{ReadingEvent}s for which the word ``pipe'' appeared in the viewport. This can be used to retrieves parts of a document which have been read (or likely to have been read) by the user. The found results can be manually filtered to retrieve only events which contained fixations, for example.

\subsubsection{search}

To find documents that contain given the text use the \texttt{search} endpoint.

\paragraph{Example}

\path{http://localhost:8080/api/eventsearch?query=pipe&type=http://www.hiit.fi/ontologies/dime/#ScientificDocument} finds all PDFs that contain the word “pipe” in any part of their body and have been previously opened by the user.

\subsection{URL types}\label{sec:urltypes}

PeyeDF also supports the native macOS ‘URL Types’ corresponding to interprocess communication protocol. That is, third party applications can ask PeyeDF to open documents and optionally focus on given areas. Is possible to send messages to PeyeDF using URLs starting with the \texttt{peyedf://} protocol. A custom URL allows for relatively simple interaction between applications.

There are two modes that can be used:

\begin{itemize}
\item \texttt{reader} opens a file for a new reading session. It can be
  used by a search engine to direct to a specific search result.
\item \texttt{refinder} opens a previously read file. Useful for personal
  indexing services which access own reading data.
\end{itemize}

These appear where the \emph{host} would normally appear in a URL (e.g.
\texttt{peyedf://refinder/\ldots})

\subsubsection{Reader}

The format of the request is:

\path{peyedf://reader/[path_on_disk|contentHash|appId|sessionId]?search=search_query[&page=n][&rect=(x,y,w,h)|&point=(x,y)]}

After \texttt{reader/} one can refer to a file using a full path. If the file is already known by DiMe, it is also possible to use a contentHash, a given sessionId or appId.

The search query is optional. If enclosed in double quotes (\texttt{"} or a \texttt{\%22} escape sequence) it is treated as an exact phrase search (as opposed to an ``all words'' search).

Optionally, one can focus to a specific block of text using the
\texttt{page} and \texttt{rect} parameters. Otherwise, \texttt{point}
and \texttt{page} can be used to focus on a broad area of text (to focus
on a section heading, for example). The \texttt{page} parameter must be
present if either \texttt{point} or \texttt{rect} is used. It is also
possible to use \texttt{page} alone, to focus on the beginning of a
given page.

\paragraph{Examples}

\path{peyedf://reader/Users/marco/Downloads/Wyble\%20et\%20al\%202009.pdf?search=\%22attentional\%20blink\%22}

This opens the `Wyble et al 2009.pdf' file and searches for the
``attentional blink'' phrase.

\path{peyedf://reader/Users/marco/Downloads/Yan\%20et\%20al\%202007.pdf?rect=(200,400,200,100)\&page=3}

This opens the `Yan et al 2007.pdf' file and focuses on the given area with origin 200 (x) , 400 (y) and size 200 (width), 100 (height).

\path{peyedf://reader/Users/marco/Downloads/Yan\%20et\%20al\%202007.pdf?rect=(200,400,200,100)&page=3}

This opens the same file as before, but instead navigates to a point
approximately in the middle of page 2 (or 1, if we start counting from 0).

\subsubsection{Refinder}

The format of the request is:

\path{peyedf://reader/Users/marco/Downloads/Yan\%20et\%20al\%202007.pdf?rect=(200,400,200,100)&page=3}

\texttt{sessionId} is the unique id assigned to each reading session, as described in \autoref{sec:events}.

Optionally, a specific block of text can be focused upon using the
\texttt{page} and \texttt{rect} parameters. Otherwise, \texttt{point}
and \texttt{page} can be used to focus on a broad area of text (to focus
on a section heading, for example). The \texttt{page} parameter must be
present if either \texttt{point} or \texttt{rect} is used. It is also
possible to use \texttt{page} alone, to focus on the beginning of a
given page.

\paragraph{Examples}

\path{peyedf://refinder/4c12e273e7be240f9eca1f?point=(200,200)&page=1}

Opens the same reading session and in addition focuses to that point on
page 2 (if starting from 1).

\path{peyedf://refinder/c7b4d50487d47ce88?rect=(100,400,200,100)&page=3}

Opens another reading session and focuses on the given block of text on
page 4 (if starting from 1). The given block of text could be something
previously annotated, for example.

\subsection{Data collection}

We, here, explain the mechanisms that are used for collecting data. We first explain how and when a \texttt{ReadingEvent} is generated and which parts of the software are involved. We then discuss the support for eye tracking devices, followed by explaining how an experiment is built using the software.

\subsubsection{ReadingEvent generation}\label{sec:readingeventgeneration}

\begin{figure*}
\centering
\includegraphics[width=\textwidth]{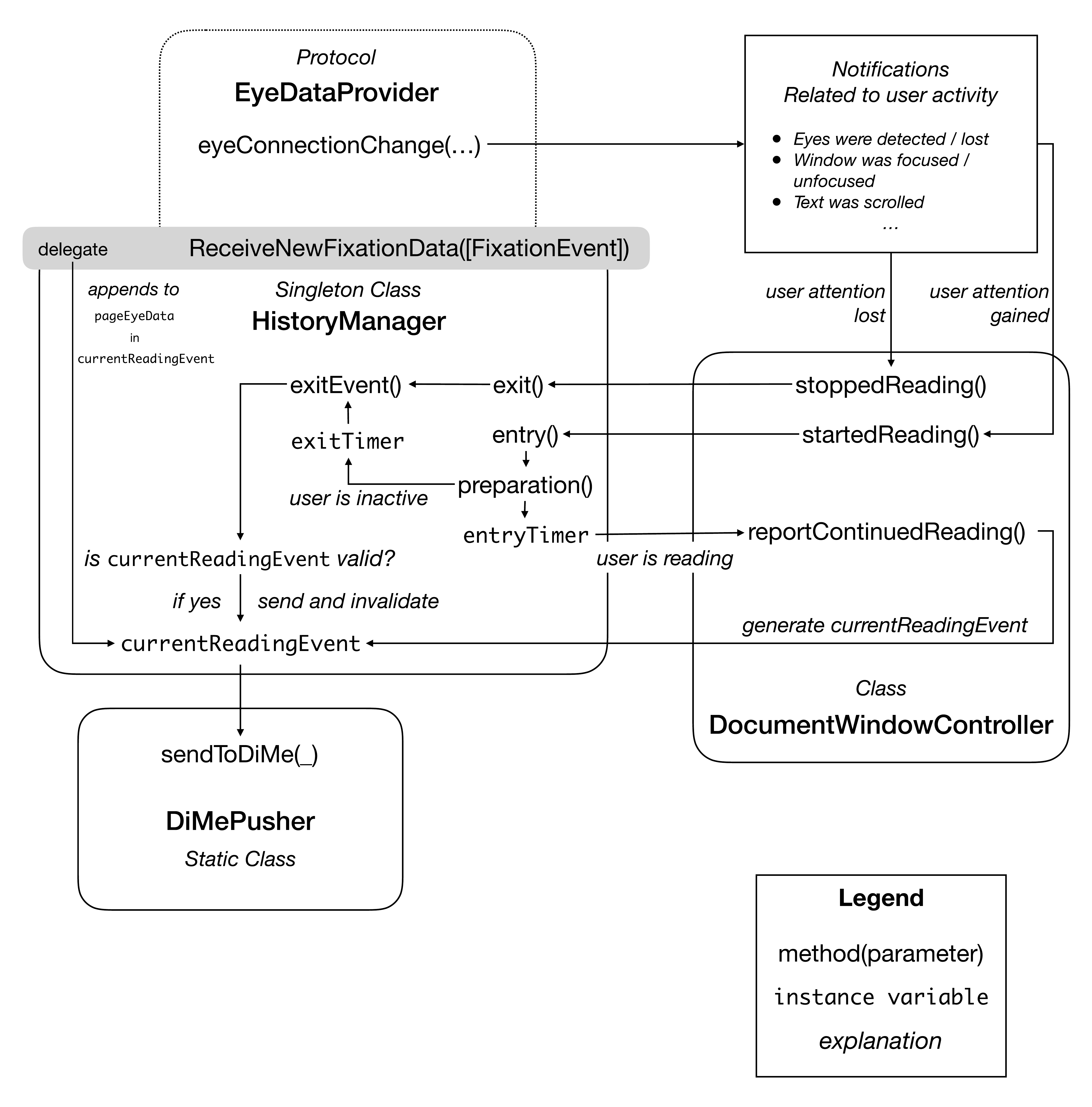}
\caption{Generation of \texttt{ReadingEvents}. This diagram depicts the flow involved in the generation of \texttt{ReadingEvents}. As detailed in \autoref{sec:performance}, this allows PeyeDF to run smoothly even when large amounts of eye tracking data are being collected and re-referenced to text and paragraphs. Events are initially dispatched via application-wide notifications (top right). These notifications are generated by events representing user or eye tracking activity. This is described in more detail in \autoref{sec:readingeventgeneration}.} \label{fig:readingeventgeneration}
\end{figure*}

The central classes involved in data collection are \texttt{HistoryManager}, \\\texttt{DocumentWindowController}, \texttt{PDFReader}, to be described below. Their interaction is visually depicted in \autoref{fig:readingeventgeneration}. 

The flow of events begins with \texttt{Notification} events, which are generated by user activity. The notifications are collected by an active \texttt{DocumentWindowController}, which triggers the \texttt{HistoryManager} by identifying itself as the centre of the user's attention. When the software detects a user has finished reading (e.g. switched to another window), the active controller notifies the \texttt{HistoryManager} again. 

Upon first call of the \texttt{HistoryManager}, it sets up a short timer (\texttt{entryTimer} 2 seconds, defined in \texttt{PeyeConstants.minReadTime}, that is keep timing meanwhile the user is reading, preventing noisy data from being recorded. The timer asks the \texttt{DocumentWindowController} to generate a \texttt{ReadingEvent}, containing the viewport information and eye tracking data when available. This event will be sent to DiMe later. The timer stops upon
 \texttt{exitTimer} which is normally triggered only when eye tracking is off and the user leaves the workstation.

\subsubsection{HistoryManager}

The \texttt{HistoryManager} class is a singleton class. That is, only one instance of it is created and is shared across the application. This class acts as a central buffer for behavioural and eye tracking data, which are frequently sent to DiMe. These data are stored in the \texttt{currentReadingEvent} field; this field is cleared and its previous contents are sent to DiMe every time exit() is called.

\subsubsection{DocumentWindowController}

 Every document window displayed by PeyeDF is an instance of \texttt{DocumentWindowController}.
 It calls the entry() and exit() methods in \texttt{HistoryManager} to record all the information.
  Upon appearing, the window sets up a number of \texttt{Notification}s, which monitor user and eye tracking activity. Every time the user “starts reading”, the DocumentWindowController calls entry() in \texttt{HistoryManager} passing itself as a parameter. This identifies the current window as the window the user is currently paying attention to. 
  When the user “stops reading” exit() is called. 
  
  The “start reading” and “stops reading” actions are defined by the type of \texttt{Notification} that are being triggered. For example, if the window \emph{starts} moving, loses focus or it is otherwise not visible it is assumed that the user stopped reading. Similarly, exit() is called when the user's gaze was lost for a relatively long amount of time (defined in \texttt{PeyeConstants.maxReadTime}). Otherwise, when the window \emph{stops} moving, or a scroll event is terminated, it is assumed that the user started reading, and entry() is called. This sets up a short timer, (\texttt{entryTimer}, 2 seconds) which upon fires calls the \texttt{HistoryManager.}entryTimerFire() method, which signals that the user is actively reading. This generates a \texttt{ReadingEvent} in the background containing PDF information (e.g. viewports), ready to be sent to DiMe once the user stops reading.

\subsubsection{PDFReader}

Every \texttt{DocumentWindowController} contains a reference to \texttt{PDFReader}. This is a subclass of \texttt{PDFBase}, which represents any instance of a PDF file displayed in PeyeDF. \texttt{PDFReader} is specialised in reading detection and interaction with the user. \texttt{PDFReader} contains the getViewPortStatus() method, called by \texttt{HistoryManager.}entryTimerFire(). \texttt{PDFReader} also contains convenience methods for creating annotations and tagging while reading. The \texttt{PDFBase} superclass contains anything relatable to a PDF, such as annotations, tags, searching support. Each \texttt{DocumentWindowController} controls one instance of \texttt{PDFReader}.

\subsection{Paragraph detection}\label{sec:paragraphdetection}

PeyeDF uses the distance of the user from the screen to determine which paragraph is currently being read (or annotated) by the user. Distance is obtained from the eye tracker, if available -- otherwise it defaults to 60~cm. When a fixation is detected and it falls within text, the \texttt{PDFBase.}pointToParagraphRect() is called by \texttt{DocumentWindowController} to determine the height covered by 3° of visual angle, which should cover abundantly the area covered by the fovea \cite{ojanpaa2002eye}. This only covers the area corresponding to the current textual paragraph; in other words, double newlines are skipped at the edges, if present (detected using textual from within the PDF information).

This information is stored in the \texttt{EyeRectangle} class, which is converted into \texttt{Rect}s and sent to DiMe within \texttt{ReadingEvent}s. When this happens, the \texttt{splitAndCrop} function is called to split \texttt{EyeRectangles} which are too big (as defined by \texttt{maxHeight}). To sum up, this procedure unites whole paragraphs unless they reach \texttt{maxHeight}. 

Algorithms and data structures regarding paragraph operations are defined in \texttt{Paragraphs.swift}\footnote{\url{https://github.com/HIIT/PeyeDF/blob/master/PeyeDF/Model/Paragraphs.swift}}. This includes uniting colliding rectangles, and converting \texttt{Rect}s stored in DiMe back to annotated blocks of text.

\subsection{Performance}\label{sec:performance}

PeyeDF was designed to maximise both usability of the software and reliability of eye tracking data. High usability is achieved by behaving similarly to a lightweight PDF reader and providing tagging and Refinder functionality. Eye tracking and behavioural data are collected using background Grand Central Dispatch\footnote{\url{https://developer.apple.com/guides/threading.html}} (GCD) queues. The main queue for eye data processing is called \texttt{eyeQueue} in \texttt{HistoryManager}. This has a default priority, which in GCD is third, below \texttt{userInteractive} and \texttt{userInitiated}. Each eye tracker implementation also has its own default priority queue (e.g. \texttt{LSLManager} has a \texttt{queue} field to which all data callbacks are dispatched).

To prevent consuming too much CPU power and reduce unnecessary data consumption, PeyeDF is conservative regarding the creation of events. It is assumed that the user is not reading unless no interaction with the software happened for a predefined amount of time (2 seconds, defined in \texttt{PeyeConstants.minReadTime}) while the window is still currently being focused (i.e. in front).

\subsection{Eye tracker support}\label{sec:generalisability}

PeyeDF currently supports three types of protocols to enable eye tracking. These are MIDAS, LSL (for SMI eye trackers) and zeroMQ (for Pupil Labs glasses).

The \texttt{Extras} folder in our repository\footnote{\url{https://github.com/HIIT/PeyeDF/tree/master/Extras}}, provides supporting files for SMI trackers and Pupil Lab glasses. For SMI, we provide a script that converts SMI network packets into LSL streams. For Pupil Labs, we provide a plugin that allows surfaces (2-D frames) to send fixations over zeroMQ. This is because PeyeDF requires fixation-level information. In case these are not available, a possible alternative is to use the MIDAS framework to calculate them in real time, which we support. This framework is described by \cite{henelius2018}.

Assuming fixation-level data can be obtained, implementing support for additional eye trackers requires two steps: 1) creating a class that adopts the \texttt{EyeDataProvider} protocol and 2) adding the given class to the \texttt{EyeDataProviderType} enum.

\subsubsection{Adopting the EyeDataProvider protocol}

The protocol generalises eye tracking behaviour by providing a number of implemented methods (protocol extensions) that should simply be called by the adopter at the correct times; these are sendLastRaw(), sendFixations(), eyeStateChange() and eyeConnectionChange(). Normally, these are be called in response to receipt of gaze data (i.e. within data callbacks, depending on the specific implementation). The protocol adopter should also provide start() and stop() methods, which control activation and deactivation of the data stream. In addition, it should hold references to the \texttt{fixationDelegate}, \texttt{available}, \texttt{eyesLost} and \texttt{fixationDelegate} fields. The provided adopters (e.g. \texttt{LSLManager}\footnote{\url{https://github.com/HIIT/PeyeDF/tree/master/PeyeDF/Eye\%20Tracking/LSLManager.swift}} can be used as a reference for new implementations.

\subsubsection{The EyeDataProviderType enum}

The enum is used a reference that informs PeyeDF on the current types of eye trackers supported. Hence, in addition to protocol implementation, a minor sequence of two-line edits to the \texttt{EyeDataProviderType} enum are required to fully integrate the eye tracker into PeyeDF. These are a \texttt{case} statement in the \texttt{EyeDataProviderType}, a new instance of the class under \texttt{associatedTracker}, a string for \texttt{description} and an increased \texttt{count} variable. For example, a tracker called \texttt{NewTracker} would be associated to a \texttt{.newtracker} enum which would call NewTracker(), a \texttt{description} of "New Tracker" while \texttt{count} should be increased from 4 to 5. The current version of \texttt{EyeDataProviderType} can be used as a reference. Completing this step will add the eye tracker to the Preferences > Experiment > Eye Tracker menu and will immediately provide eye tracking data upon selection.

\subsection{Experiment creation}\label{sec:experimentcreation}

PeyeDF can be utilised to conduct user experiments. The approach we recommend is to create a git branch (in github, or a private local branch). Subsequently it is suggested to create an additional target that will be used to run the experiment. It is possible to duplicate the original ‘PeyeDF’ target into another (e.g. ‘PeyeDF Experiment 1’).

Under the Xcode target's build settings, additional flags can be added to identify the target in the application's code. That is, under ‘Other Swift Flags’ create a flag using the \texttt{-Dkeyword} syntax. For example, a flag called \texttt{ExperimentalTarget} requires a \texttt{-DExperimentalTarget} flag. One can then add target-specific code using:

\begin{lstlisting}[language=Elan]
#if ExperimentalTarget
    // anything experiment-specific
#endif
\end{lstlisting}

This way, patches can be “pushed upstream” or “pulled downstream” independently of experiment-specific implementations. We provide the ‘PeyeDF Questions’ target as an example of how to define experimental targets. This target starts the Refinder and will run through a predefined list of papers, asking a given set of multiple-choice questions. It was utilised to run the second session of the experiment described in the following section.

\section{Experiment}\label{sec:experiment}

The experiment tested the possibility of utilising fixation data collected in the past to predict reading comprehension which takes place in the future. That is, fixation data collected during a first reading of a given passage was utilised to predict performance (answer correctness) that took place during a subsequent reading of the same passage. The experiment was ran utilising PeyeDF, while data analysis was performed in Matlab 2018a.

The study was divided into two sessions. During the first session, participants were instructed to read a sequence of papers and find specific information within each. The second session took place exactly one week after the first; participants revisited the papers, answering multiple-choice questions that referred to the information they had to locate during the first session. We demonstrate that fixation data collected during the first session predicts performance in the second.

\subsection{Question creation}\label{sec:questioncreation}

We selected four papers, and in relation to each we created a number of questions that referred to a specific passage. We selected papers with the following requirements: they had to be relatively easy to read, while covering topics of general interest (e.g. health). They also had to contain enough text and information to allow the creation of a total of 24 questions. The questions also had to be grouped so that 4 consecutive questions were related to the same topic (and were therefore located within the same section / paragraph). We gave the name ``target topic'' to the topic covered by a sequence of four related questions. This way, we obtained 6 ``target topics'' per paper. We then split the 6 target topics into two groups of three. The resulting two ``target topic groups'' were called A and B. The split was done randomly but did not change across participants (e.g. group A in paper 1 always referred to the same set of questions).

To summarise, each target topic group (A or B) contained 3 target topics, each containing 4 questions. One participant would then be assigned only one target topic group within a paper. Splitting into target topic groups allowed us to assign a different set of questions to every participant, even though they all read the same papers.

\begin{samepage}
The four papers we utilised were:
\begin{enumerate}
    \item The Impact of Asthma and Allergic Diseases on Schoolchildren: Are They at Increased Risk of Absenteeism and Poor School Performance? \cite{bener2011impact}
    \item Acceptability of psychotherapy, pharmacotherapy, and self-directed therapies in Australians living with chronic hepatitis C \cite{stewart2013acceptability}
    \item The effectiveness of a motivational interviewing primary-care based intervention on physical activity and predictors of change in a disadvantaged community \cite{hardcastle2012effectiveness}
    \item Choice and placebo expectation effects in the context of pain analgesia \cite{rose2012choice}
\end{enumerate}
\end{samepage}

An additional paper (titled Waterpipe smoking among US university students, \cite{primack2012waterpipe}) was used for practice. This contained only 2 topics with 2 questions each in total, all belonging to the same group (group A).

The resulting set of questions were saved in JSON files and are available in our GitHub repository\footnote{\url{https://github.com/HIIT/PeyeDF/tree/master/Extras/Experiment/Questions}\label{ft:questions}}.

We used PeyeDF’s tagging functionality to identify the location in the text corresponding to the answer to each question. This information was stored in JSON files, available in our repository\footnote{\url{https://github.com/HIIT/PeyeDF/tree/master/Extras/Experiment/Answer\_Location\_Tags}\label{ft:answerlocationtags}}. Each JSON file contains the answer location for a given paper (1 to 4), group (A or B) and target topic (1 to 3). The answer location corresponds to a page and a rectangle (text box) which encloses the text containing the answer (along with the actual text of the answer).

\subsubsection{First session}

Each participant was assigned the set of four papers, randomly shuffled and with randomly assigned target topic groups. An assignment could then be represented as 1A, 2B, 4B, 3B, indicating that the participant had to start from paper 1 and would only be shown questions from group A, then continue to paper 2, group B, etc.

Participants were given 15 minutes to freely familiarise themselves with each paper before being asked to perform the information finding task. After this familiarisation phase, participants were asked to find specific information by showing to them instructions on a computer screen. The following example depicts instructions for paper 1, group A, target topics 1 and 2:

\begin{itemize}
    \item In Bener2011\_asthma.pdf
    \begin{itemize}
        \item An overall (worldwide) view of asthma \emph{(target topic 1)}
        \begin{itemize}
            \item Overall asthma prevalence
            \item Factors correlated to asthma
            \item Lifestyle disease modifier factors
            \item Number of people affected by pollutants
        \end{itemize}
    \item Findings about asthma and allergic diseases \emph{(target topic 2)}
        \begin{itemize}
            \item Most accepted method to measure asthma prevalence
            \item Correlation between age and pulmonary infection
            \item Relationship to sex
            \item Relationship between rhinitis prevalence and sex
        \end{itemize}
    \end{itemize}
\end{itemize}

The text shown for each item (identified by asterisk in the example) is called ``summary'' and is saved in the JSON files available in our repository$^{\ref{ft:questions}}$.

We utilised the practice paper to guide participants in this task, ensuring they understood the procedure correctly.

Participants were asked to simply look up the information, without putting any additional effort in memorising it. Participants were instructed to proceed to the next paper after finding all the necessary information. They were also asked to advise the experimenter immediately in case they found it difficult to locate the information they were asked to find (pilot testing was used to verify that all information could be found autonomously by participants). Once finished, they were instructed to advise the experimenter.

\subsubsection{Second session}

The second session was run using PeyeDF’s Refinder functionality (previously shown in \autoref{fig:refinderscreenshot}). This feature automatically displayed the papers the given participant read during the first session, in order. The Refinder funcionality was modified using the ‘PeyeDF Questions’ target\footnote{\url{https://github.com/HIIT/PeyeDF/tree/master/PeyeDF/Questions}}. This modification displays a multiple choice question below the Refinder. Participants had to select one out of three possible answers and confirm their answer before proceeding to the next question. They performed this task in the same order as session one.

\begin{samepage}
The following example refers to paper 1, group A, target topic 1, question 1:
\begin{itemize}
    \item In western countries, how high can the prevalence of asthma and allergies be (among children)?
    \begin{itemize}
        \item 1/3
        \item 1/4
        \item 1/5
    \end{itemize}
\end{itemize}
\end{samepage}

The first item of the Question JSON files$^{\ref{ft:questions}}$ always corresponds to the correct answer. Answers were displayed to participants in random order. Participants were trained for this task using the practice paper, along with the experimenter.

Note that questions do not exactly match the ‘summary’ participants were shown during the first session. Our aim was to induce participants into re-reading the text, rather than simply rely on their memory. This is because we were interested in assessing whether fixation data collected in the past predicts reading comprehension in the future (correlations between fixation duration and memory have already been demonstrated in \cite{hollingworth2001see}, for example).

Response correctness and time passed in relation to each question were recorded and exported to JSON for analysis. 

\subsection{Participants}

In total, we recruited 12 participants. We rejected participants that spent less than 30 minutes or more than 90 minutes to complete the first session. We also rejected participants when eye tracking calibration error was above 1° of visual angle. This resulted in 7 valid participants. We also set a threshold for rejection at 80~\% correct answers given in the second session; however, none of the remaining participants were below this threshold.

Each participant was paid with two movie tickets.

\subsection{Data analysis}

As mentioned in \autoref{sec:questioncreation}, we used Tags to store the location within the text (page number and text box) corresponding to each answer. We utilised these to identify all fixations collected in the first session that fell within 3° of visual angle from the centre of a given answer location. This way, we obtained a set of fixation positions and durations in relation to each answer, for each participant. No fixation data from the second session was utilised.

We populated a Matlab table in which each row corresponded to an answer. Each column represented information related to the answer (correctness, time spent to answer) and the question (target topic number, group, paper number). Each column also contained features calculated from eye tracking data such as average fixation duration and travel. We created three Support Vector Machine classifiers that predicted response correctness using a given set of columns; the classifiers and related features (table columns) were:

\begin{itemize}
    \item \textbf{Eye}: classifier for gaze data
    \begin{itemize}
        \item Average fixation duration
        \item Median fixation duration
        \item Sum of fixation durations
        \item Distance of eyes from screen
        \item Total travel (sum of saccades)
        \item Forward travel (sum of saccades moving left or towards the bottom of the screen)
        \item Backward travel (sum of saccades moving upwards or towards the top of the screen)
    \end{itemize}
    \item \textbf{Topic}: classifier for question data
    \begin{itemize}
        \item Paper number (1 to 4)
        \item Group identity (A or B)
        \item Target topic number (1 to 3)
    \end{itemize}
    \item \textbf{All}: classifier that considered all table columns.
\end{itemize}

All classifiers were trained using leave-one-out cross validation. We utilised the Area Under the ROC Curve (AUC) as the performance measure for each classifier, as this has shown to be appropriate for unbalanced data data~\cite{jeni2013facing}. Our data was unbalanced since the predicted variable (response correctness) was skewed, since 87\% of responses were correct.

We utilised a permutation test \cite{Good2013} to compute a p-value for each classifier. That is, every classifier was run an additional 1000 times with the predicted variable (response correctness) randomly permuted across table rows. A p-value is obtained by dividing the number of times that a randomised AUC was greater than the true-observed AUC by the number of permutations.

\subsection{Results and discussion}

Of the tree classifiers, only \textbf{All} was significant (AUC: .62, \emph{p} = 0.041). \textbf{Topic} (AUC: .37, \emph{p} > 0.5) and \textbf{Eye} (AUC: .50, \emph{p} > 0.5) failed to classify response correctness above chance levels. This indicates that gaze data significantly contributed to predictions only when coupled to question-related information. However, question related information alone (the Topic classifier) was not sufficient to predict response correctness. Similarly, gaze data alone is not sufficient. This indicates collecting gaze data for future use, which the main feature of PeyeDF, is valuable for reading-based experiments, as long as some text-related information is also collected. We believe this is because not all target topics were of identical difficulty. Combining topic data with gaze-related data provides enough information to discriminate between topics which were been previously read carefully and were relatively easy to understand, agains topics which were not carefully read and / or were harder to understand.

Note that topic data was linked to the design of the experiment in this example. In more realistic scenarios, topic features could be replaced with natural language metrics (such as number nouns and verbs) \cite{bird2009natural} or approximate location (page number, section) of the text currently read by the user.

During pilot testing, SMI eye tracker reported a pupil size of 0, indicating that pupil size could not be estimated reliably (this could be due to lighting in our lab or the fact that we used a highly interactive application). Gaze-based classification could be further improved by including pupil size measurements, as these correlate to workload \cite{beatty2000pupillary}. Despite this limitation, we demonstrated that fixation data provides valuable information and can significantly help in discriminating between correctly and incorrectly comprehended text.

New frameworks recently released for the macOS platform such as the Basic neural network subroutines (BNSS - part of the Accelerate framework), the Convolutional Neural Networks (CNN - part of the MetalPerformanceShaders framework) or the upcoming Create ML framework could be integrated in the future, making deployment of machine learning applications and experiments integrated within PeyeDF, improving user-friendliness.

Collaborative functions are included in PeyeDF. The experimental design we utilised in the previous section could be adapted for collaborative reading research. In particular, we identified two independent ``target topic groups'' within a number of papers. Each paper could then be read by two participants, one focusing on group A and the other on group B. Gaze and behavioural data could then be used to investigate wether their features can distinguish between the two participants, for example.

PeyeDF is a desktop application; however, many of its functions are based on PDFKit, which is a framework recently made available for iOS. This means that PeyeDF could be extended to run on tablets or smartphones, given compatible eye-tracking hardware. It could also run as an eye-tracking data viewer only.

PeyeDF enables longer term research than traditional experimental software. That is, machine shutdowns and startups do not interfere with the data, which is stored in DiMe. Hence, it can be used to track how user habits change over time using its integration of behavioural and eye tracking data stored within \texttt{ReadingEvent}s. For example, it has been suggested that expertise affects gaze over time, on the same participants \cite{vangog2010}. PeyeDF would be suitable for experiments in this directions, as experiments can be run over long periods of time. In educational settings, PeyeDF could be used to assess retention of read material depending on eye tracking data \cite{rayner2006eye}.

The Refinder functionality included in PeyeDF is relatively basic and could be extended to investigate how different implementation affect information retrieval and retention. For example, it would be possible to obtain read paragraph information from \texttt{ReadingEvent}s and use it to compute a cloud of potentially related documents as suggested by \cite{Buscher2012,buscher2009gaze}.

MIDAS \cite{henelius2018}, which is supported by PeyeDF, allows real-time processing of sensor data. It could be used to gather data collected from additional sensors, enabling research on the relationship between physiology and emotions experienced during reading, as suggested by \cite{graesser2012moment}.

Native annotation support in PeyeDF is useful for annotation-directed research. That is, different types of annotation can be proposed to users; subsequent DiMe queries can be used to determine which were the most used annotations and experiments can be designed to assess information retention depending on annotation type.

PeyeDF can also be used for more controlled, short-term studies. PDF pages contained the required stimuli can be created and read by PDF. Full screen support allows to display the required stimuli covering the whole screen, if required. Page movements could then be triggered by code, rather than user activity.

\section*{Conflict of Interest Statement}

The authors declare that the research was conducted in the absence of any commercial or financial relationships that could be construed as a potential conflict of interest.

\section*{Author Contributions}

MF wrote the paper, designed the software and the related experiment. HT contributed to the writing of the article and to the analysis of eye tracking data. NR and GJ supervised the software and experimental design.

\section*{Funding}
The present work has been supported by the Finnish Funding Agency for Innovation (TEKES) under the project Revolution of Knowledge Work (funding decisions 558/31/2013 and 5159/31/2014).

\bibliographystyle{ACM-Reference-Format}
\bibliography{peyedf}


\begin{thebibliography}{70}


\ifx \showCODEN    \undefined \def \showCODEN     #1{\unskip}     \fi
\ifx \showDOI      \undefined \def \showDOI       #1{#1}\fi
\ifx \showISBNx    \undefined \def \showISBNx     #1{\unskip}     \fi
\ifx \showISBNxiii \undefined \def \showISBNxiii  #1{\unskip}     \fi
\ifx \showISSN     \undefined \def \showISSN      #1{\unskip}     \fi
\ifx \showLCCN     \undefined \def \showLCCN      #1{\unskip}     \fi
\ifx \shownote     \undefined \def \shownote      #1{#1}          \fi
\ifx \showarticletitle \undefined \def \showarticletitle #1{#1}   \fi
\ifx \showURL      \undefined \def \showURL       {\relax}        \fi
\providecommand\bibfield[2]{#2}
\providecommand\bibinfo[2]{#2}
\providecommand\natexlab[1]{#1}
\providecommand\showeprint[2][]{arXiv:#2}

\bibitem[\protect\citeauthoryear{Ackerman and Goldsmith}{Ackerman and
  Goldsmith}{2011}]%
        {ackerman2011}
\bibfield{author}{\bibinfo{person}{R Ackerman} {and} \bibinfo{person}{M
  Goldsmith}.} \bibinfo{year}{2011}\natexlab{}.
\newblock \showarticletitle{Metacognitive regulation of text learning: on
  screen versus on paper.}
\newblock \bibinfo{journal}{\emph{J Exp Psychol Appl}} \bibinfo{volume}{17},
  \bibinfo{number}{1} (\bibinfo{year}{2011}), \bibinfo{pages}{18--32}.
\newblock


\bibitem[\protect\citeauthoryear{Adler, Gujar, Harrison, O'Hara, and
  Sellen}{Adler et~al\mbox{.}}{1998}]%
        {adler1998diary}
\bibfield{author}{\bibinfo{person}{Annette Adler}, \bibinfo{person}{Anuj
  Gujar}, \bibinfo{person}{Beverly~L Harrison}, \bibinfo{person}{Kenton
  O'Hara}, {and} \bibinfo{person}{Abigail Sellen}.}
  \bibinfo{year}{1998}\natexlab{}.
\newblock \showarticletitle{A diary study of work-related reading: design
  implications for digital reading devices Proceedings of the {SIGCHI}
  conference on Human factors in computing systems}. \bibinfo{publisher}{ACM
  Press/Addison-Wesley Publishing Co.}, \bibinfo{pages}{241--248}.
\newblock


\bibitem[\protect\citeauthoryear{Alamargot, Chesnet, Dansac, and Ros}{Alamargot
  et~al\mbox{.}}{2006}]%
        {Alamargot:2006aa}
\bibfield{author}{\bibinfo{person}{D Alamargot}, \bibinfo{person}{D Chesnet},
  \bibinfo{person}{C Dansac}, {and} \bibinfo{person}{C Ros}.}
  \bibinfo{year}{2006}\natexlab{}.
\newblock \showarticletitle{Eye and pen: A new device for studying reading
  during writing.}
\newblock \bibinfo{journal}{\emph{Behavior Research Methods}}
  \bibinfo{volume}{38}, \bibinfo{number}{2} (\bibinfo{year}{2006}),
  \bibinfo{pages}{287--299}.
\newblock


\bibitem[\protect\citeauthoryear{Ball and Hourcade}{Ball and Hourcade}{2011}]%
        {ball2011rethinking}
\bibfield{author}{\bibinfo{person}{Robert Ball} {and}
  \bibinfo{person}{Juan~Pablo Hourcade}.} \bibinfo{year}{2011}\natexlab{}.
\newblock \showarticletitle{Rethinking Reading for Age From Paper and
  Computers}.
\newblock \bibinfo{journal}{\emph{International Journal of Human-Computer
  Interaction}} \bibinfo{volume}{27}, \bibinfo{number}{11}
  (\bibinfo{year}{2011}), \bibinfo{pages}{1066--1082}.
\newblock


\bibitem[\protect\citeauthoryear{Beatty, Lucero-Wagoner, et~al\mbox{.}}{Beatty
  et~al\mbox{.}}{2000}]%
        {beatty2000pupillary}
\bibfield{author}{\bibinfo{person}{Jackson Beatty}, \bibinfo{person}{Brennis
  Lucero-Wagoner}, {et~al\mbox{.}}} \bibinfo{year}{2000}\natexlab{}.
\newblock \showarticletitle{The pupillary system}.
\newblock \bibinfo{journal}{\emph{Handbook of psychophysiology}}
  \bibinfo{volume}{2}, \bibinfo{number}{142-162} (\bibinfo{year}{2000}).
\newblock


\bibitem[\protect\citeauthoryear{Ben-Yehudah and Eshet-Alkalai}{Ben-Yehudah and
  Eshet-Alkalai}{2014}]%
        {benyehudah2014influence}
\bibfield{author}{\bibinfo{person}{Gal Ben-Yehudah} {and}
  \bibinfo{person}{Yoram Eshet-Alkalai}.} \bibinfo{year}{2014}\natexlab{}.
\newblock \showarticletitle{The Influence of Text Annotation Tools on Print and
  Digital Reading Comprehension}. In \bibinfo{booktitle}{\emph{Proceedings of
  the 9th Chais Conference for the Study of Innovation and Learning
  Technologies: Learning in the Technological Era}},
  \bibfield{editor}{\bibinfo{person}{Y.~Eshet-Alkalai},
  \bibinfo{person}{A.~Caspi}, \bibinfo{person}{N.~Geri},
  \bibinfo{person}{Y.~Kalman}, \bibinfo{person}{V.~Silber-Varod}, {and}
  \bibinfo{person}{Y.~Yair}} (Eds.). \bibinfo{publisher}{The Open University of
  Israel}, \bibinfo{pages}{28--35}.
\newblock


\bibitem[\protect\citeauthoryear{Bener}{Bener}{2011}]%
        {bener2011impact}
\bibfield{author}{\bibinfo{person}{Abdulbari Bener}.}
  \bibinfo{year}{2011}\natexlab{}.
\newblock \showarticletitle{The Impact of Asthma and Allergic Diseases on
  Schoolchildren: Are They at Increased Risk of Absenteeism and Poor School
  Performance?}
\newblock In \bibinfo{booktitle}{\emph{Advanced Topics in Environmental Health
  and Air Pollution Case Studies}}. \bibinfo{publisher}{InTech}.
\newblock


\bibitem[\protect\citeauthoryear{Biedert, Buscher, and Dengel}{Biedert
  et~al\mbox{.}}{2010a}]%
        {Biedert2010}
\bibfield{author}{\bibinfo{person}{Ralf Biedert}, \bibinfo{person}{Georg
  Buscher}, {and} \bibinfo{person}{Andreas Dengel}.}
  \bibinfo{year}{2010}\natexlab{a}.
\newblock \showarticletitle{The eyeBook -- Using Eye Tracking to Enhance the
  Reading Experience}.
\newblock \bibinfo{journal}{\emph{Informatik Spektrum}} \bibinfo{volume}{33},
  \bibinfo{number}{3} (\bibinfo{year}{2010}), \bibinfo{pages}{272--281}.
\newblock


\bibitem[\protect\citeauthoryear{Biedert, Buscher, Schwarz, M{\"o}ller, Dengel,
  and Lottermann}{Biedert et~al\mbox{.}}{2010b}]%
        {biedert2010text}
\bibfield{author}{\bibinfo{person}{Ralf Biedert}, \bibinfo{person}{Georg
  Buscher}, \bibinfo{person}{Sven Schwarz}, \bibinfo{person}{Manuel
  M{\"o}ller}, \bibinfo{person}{Andreas Dengel}, {and} \bibinfo{person}{Thomas
  Lottermann}.} \bibinfo{year}{2010}\natexlab{b}.
\newblock \showarticletitle{The text 2.0 framework}. In
  \bibinfo{booktitle}{\emph{Workshop on Eye Gaze in Intelligent Human Machine
  Interaction}}. \bibinfo{publisher}{Citeseer}, \bibinfo{pages}{114--117}.
\newblock


\bibitem[\protect\citeauthoryear{Bird, Klein, and Loper}{Bird
  et~al\mbox{.}}{2009}]%
        {bird2009natural}
\bibfield{author}{\bibinfo{person}{Steven Bird}, \bibinfo{person}{Ewan Klein},
  {and} \bibinfo{person}{Edward Loper}.} \bibinfo{year}{2009}\natexlab{}.
\newblock \bibinfo{booktitle}{\emph{Natural language processing with Python:
  analyzing text with the natural language toolkit}}.
\newblock \bibinfo{publisher}{O'Reilly Media}.
\newblock


\bibitem[\protect\citeauthoryear{Blair, Watson, Walshe, and Maj}{Blair
  et~al\mbox{.}}{2009}]%
        {Blair2009}
\bibfield{author}{\bibinfo{person}{MR Blair}, \bibinfo{person}{MR Watson},
  \bibinfo{person}{RC Walshe}, {and} \bibinfo{person}{F Maj}.}
  \bibinfo{year}{2009}\natexlab{}.
\newblock \showarticletitle{Extremely selective attention: eye-tracking studies
  of the dynamic allocation of attention to stimulus features in
  categorization.}
\newblock \bibinfo{journal}{\emph{Journal of experimental psychology. Learning,
  memory, and cognition}} \bibinfo{volume}{35}, \bibinfo{number}{5}
  (\bibinfo{year}{2009}), \bibinfo{pages}{1196--1206}.
\newblock


\bibitem[\protect\citeauthoryear{Borji and Itti}{Borji and Itti}{2014}]%
        {borji2014}
\bibfield{author}{\bibinfo{person}{Ali Borji} {and} \bibinfo{person}{Laurent
  Itti}.} \bibinfo{year}{2014}\natexlab{}.
\newblock \showarticletitle{Defending Yarbus: Eye movements reveal observers'
  task}.
\newblock \bibinfo{journal}{\emph{Journal of Vision}} \bibinfo{volume}{14},
  \bibinfo{number}{3} (\bibinfo{year}{2014}), \bibinfo{pages}{29}.
\newblock
\urldef\tempurl%
\url{https://doi.org/10.1167/14.3.29}
\showDOI{\tempurl}
\showeprint{/data/journals/jov/932817/i1534-7362-14-3-29.pdf}


\bibitem[\protect\citeauthoryear{Brennan, Chen, Dickinson, Neider, and
  Zelinsky}{Brennan et~al\mbox{.}}{2008}]%
        {brennan2008}
\bibfield{author}{\bibinfo{person}{SE Brennan}, \bibinfo{person}{X Chen},
  \bibinfo{person}{CA Dickinson}, \bibinfo{person}{MB Neider}, {and}
  \bibinfo{person}{GJ Zelinsky}.} \bibinfo{year}{2008}\natexlab{}.
\newblock \showarticletitle{Coordinating cognition: the costs and benefits of
  shared gaze during collaborative search.}
\newblock \bibinfo{journal}{\emph{Cognition}} \bibinfo{volume}{106},
  \bibinfo{number}{3} (\bibinfo{year}{2008}), \bibinfo{pages}{1465--1477}.
\newblock


\bibitem[\protect\citeauthoryear{Buscher and Dengel}{Buscher and
  Dengel}{2009}]%
        {buscher2009gaze}
\bibfield{author}{\bibinfo{person}{Georg Buscher} {and}
  \bibinfo{person}{Andreas Dengel}.} \bibinfo{year}{2009}\natexlab{}.
\newblock \showarticletitle{Gaze-based filtering of relevant document
  segments}. In \bibinfo{booktitle}{\emph{International World Wide Web
  Conference (WWW)}}. \bibinfo{pages}{20--24}.
\newblock


\bibitem[\protect\citeauthoryear{Buscher, Dengel, Biedert, and Elst}{Buscher
  et~al\mbox{.}}{2012}]%
        {Buscher2012}
\bibfield{author}{\bibinfo{person}{Georg Buscher}, \bibinfo{person}{Andreas
  Dengel}, \bibinfo{person}{Ralf Biedert}, {and} \bibinfo{person}{Ludger~V.
  Elst}.} \bibinfo{year}{2012}\natexlab{}.
\newblock \showarticletitle{Attentive documents: Eye tracking as implicit
  feedback for information retrieval and beyond}.
\newblock \bibinfo{journal}{\emph{ACM Trans. Interact. Intell. Syst.}}
  \bibinfo{volume}{1}, \bibinfo{number}{2} (\bibinfo{year}{2012}),
  \bibinfo{pages}{1--30}.
\newblock


\bibitem[\protect\citeauthoryear{Campbell and Maglio}{Campbell and
  Maglio}{2001}]%
        {campbell2001robust}
\bibfield{author}{\bibinfo{person}{Christopher~S Campbell} {and}
  \bibinfo{person}{Paul~P Maglio}.} \bibinfo{year}{2001}\natexlab{}.
\newblock \showarticletitle{A robust algorithm for reading detection}. In
  \bibinfo{booktitle}{\emph{Proceedings of the 2001 workshop on Perceptive user
  interfaces}}. \bibinfo{publisher}{ACM}, \bibinfo{pages}{1--7}.
\newblock


\bibitem[\protect\citeauthoryear{Chen, Cheng, Chang, Zheng, and Huang}{Chen
  et~al\mbox{.}}{2014}]%
        {chen2014}
\bibfield{author}{\bibinfo{person}{Guang Chen}, \bibinfo{person}{Wei Cheng},
  \bibinfo{person}{Ting-Wen Chang}, \bibinfo{person}{Xiaoxia Zheng}, {and}
  \bibinfo{person}{Ronghuai Huang}.} \bibinfo{year}{2014}\natexlab{}.
\newblock \showarticletitle{A comparison of reading comprehension across paper,
  computer screens, and tablets: Does tablet familiarity matter}.
\newblock \bibinfo{journal}{\emph{J. Comput. Educ.}} \bibinfo{volume}{1},
  \bibinfo{number}{2-3} (\bibinfo{year}{2014}), \bibinfo{pages}{213--225}.
\newblock


\bibitem[\protect\citeauthoryear{Cherubini, N{\"u}ssli, and
  Dillenbourg}{Cherubini et~al\mbox{.}}{2010}]%
        {cherubini2010indicating}
\bibfield{author}{\bibinfo{person}{Mauro Cherubini},
  \bibinfo{person}{Marc-Antoine N{\"u}ssli}, {and} \bibinfo{person}{Pierre
  Dillenbourg}.} \bibinfo{year}{2010}\natexlab{}.
\newblock \showarticletitle{This is it!: Indicating and looking in
  collaborative work at distance}.
\newblock \bibinfo{journal}{\emph{Journal of Eye Movement Research}}
  \bibinfo{volume}{3} (\bibinfo{year}{2010}), \bibinfo{pages}{1--20}.
\newblock


\bibitem[\protect\citeauthoryear{Djamasbi, Siegel, and Tullis}{Djamasbi
  et~al\mbox{.}}{2010}]%
        {djamasbi2010}
\bibfield{author}{\bibinfo{person}{Soussan Djamasbi}, \bibinfo{person}{Marisa
  Siegel}, {and} \bibinfo{person}{Tom Tullis}.}
  \bibinfo{year}{2010}\natexlab{}.
\newblock \showarticletitle{Generation Y, web design, and eye tracking}.
\newblock \bibinfo{journal}{\emph{International Journal of Human-Computer
  Studies}} \bibinfo{volume}{68}, \bibinfo{number}{5} (\bibinfo{year}{2010}),
  \bibinfo{pages}{307--323}.
\newblock


\bibitem[\protect\citeauthoryear{Dourish, Edwards, LaMarca, and
  Salisbury}{Dourish et~al\mbox{.}}{1999}]%
        {dourish1999presto}
\bibfield{author}{\bibinfo{person}{Paul Dourish}, \bibinfo{person}{W~Keith
  Edwards}, \bibinfo{person}{Anthony LaMarca}, {and} \bibinfo{person}{Michael
  Salisbury}.} \bibinfo{year}{1999}\natexlab{}.
\newblock \showarticletitle{Presto: an experimental architecture for fluid
  interactive document spaces}.
\newblock \bibinfo{journal}{\emph{ACM Transactions on Computer-Human
  Interaction (TOCHI)}} \bibinfo{volume}{6}, \bibinfo{number}{2}
  (\bibinfo{year}{1999}), \bibinfo{pages}{133--161}.
\newblock


\bibitem[\protect\citeauthoryear{Dragunov, Dietterich, Johnsrude, McLaughlin,
  Li, and Herlocker}{Dragunov et~al\mbox{.}}{2005}]%
        {dragunov2005tasktracer}
\bibfield{author}{\bibinfo{person}{Anton~N Dragunov}, \bibinfo{person}{Thomas~G
  Dietterich}, \bibinfo{person}{Kevin Johnsrude}, \bibinfo{person}{Matthew
  McLaughlin}, \bibinfo{person}{Lida Li}, {and} \bibinfo{person}{Jonathan~L
  Herlocker}.} \bibinfo{year}{2005}\natexlab{}.
\newblock \showarticletitle{TaskTracer: a desktop environment to support
  multi-tasking knowledge workers Proceedings of the 10th international
  conference on Intelligent user interfaces}. \bibinfo{publisher}{ACM},
  \bibinfo{pages}{75--82}.
\newblock


\bibitem[\protect\citeauthoryear{Duchowski}{Duchowski}{2002}]%
        {duchowski2002}
\bibfield{author}{\bibinfo{person}{AT Duchowski}.}
  \bibinfo{year}{2002}\natexlab{}.
\newblock \showarticletitle{A breadth-first survey of eye-tracking
  applications.}
\newblock \bibinfo{journal}{\emph{Behav Res Methods Instrum Comput}}
  \bibinfo{volume}{34}, \bibinfo{number}{4} (\bibinfo{year}{2002}),
  \bibinfo{pages}{455--470}.
\newblock


\bibitem[\protect\citeauthoryear{Dumais, Cutrell, Cadiz, Jancke, Sarin, and
  Robbins}{Dumais et~al\mbox{.}}{2016}]%
        {dumais2016stuff}
\bibfield{author}{\bibinfo{person}{Susan Dumais}, \bibinfo{person}{Edward
  Cutrell}, \bibinfo{person}{Jonathan~J Cadiz}, \bibinfo{person}{Gavin Jancke},
  \bibinfo{person}{Raman Sarin}, {and} \bibinfo{person}{Daniel~C Robbins}.}
  \bibinfo{year}{2016}\natexlab{}.
\newblock \showarticletitle{Stuff I've seen: a system for personal information
  retrieval and re-use {ACM} {SIGIR} Forum}, Vol.~\bibinfo{volume}{49(2)}.
  \bibinfo{publisher}{ACM}, \bibinfo{pages}{28--35}.
\newblock


\bibitem[\protect\citeauthoryear{Fowler and Barker}{Fowler and Barker}{1974}]%
        {Fowler1974}
\bibfield{author}{\bibinfo{person}{Robert~L Fowler} {and}
  \bibinfo{person}{Anne~S Barker}.} \bibinfo{year}{1974}\natexlab{}.
\newblock \showarticletitle{Effectiveness of highlighting for retention of text
  material.}
\newblock \bibinfo{journal}{\emph{Journal of Applied Psychology}}
  \bibinfo{volume}{59}, \bibinfo{number}{3} (\bibinfo{year}{1974}),
  \bibinfo{pages}{358}.
\newblock


\bibitem[\protect\citeauthoryear{Franze, Marriott, and Wybrow}{Franze
  et~al\mbox{.}}{2014}]%
        {franze2014}
\bibfield{author}{\bibinfo{person}{Juliane Franze}, \bibinfo{person}{Kim
  Marriott}, {and} \bibinfo{person}{Michael Wybrow}.}
  \bibinfo{year}{2014}\natexlab{}.
\newblock \showarticletitle{What academics want when reading digitally},
  \bibfield{editor}{\bibinfo{person}{Steven Simske} {and}
  \bibinfo{person}{Sebastian R{\"o}nnau}} (Eds.), Vol.~\bibinfo{volume}{the
  2014 ACM symposium}. \bibinfo{publisher}{ACM Press}, \bibinfo{address}{New
  York, New York, USA}, \bibinfo{pages}{199--202}.
\newblock


\bibitem[\protect\citeauthoryear{Good}{Good}{2013}]%
        {Good2013}
\bibfield{author}{\bibinfo{person}{Phillip Good}.}
  \bibinfo{year}{2013}\natexlab{}.
\newblock \bibinfo{booktitle}{\emph{Permutation tests: a practical guide to
  resampling methods for testing hypotheses}}.
\newblock \bibinfo{publisher}{Springer Science \& Business Media}.
\newblock


\bibitem[\protect\citeauthoryear{Graesser and D'Mello}{Graesser and
  D'Mello}{2012}]%
        {graesser2012moment}
\bibfield{author}{\bibinfo{person}{Arthur~C. Graesser} {and}
  \bibinfo{person}{Sidney D'Mello}.} \bibinfo{year}{2012}\natexlab{}.
\newblock \showarticletitle{Moment-To-Moment Emotions During Reading}.
\newblock \bibinfo{journal}{\emph{Read Teach}} \bibinfo{volume}{66},
  \bibinfo{number}{3} (\bibinfo{year}{2012}), \bibinfo{pages}{238--242}.
\newblock


\bibitem[\protect\citeauthoryear{Gurrin, Smeaton, and Doherty}{Gurrin
  et~al\mbox{.}}{2014}]%
        {gurrin2014}
\bibfield{author}{\bibinfo{person}{Cathal Gurrin}, \bibinfo{person}{Alan~F
  Smeaton}, {and} \bibinfo{person}{Aiden~R Doherty}.}
  \bibinfo{year}{2014}\natexlab{}.
\newblock \showarticletitle{Lifelogging: Personal big data}.
\newblock \bibinfo{journal}{\emph{Foundations and Trends(R) in Information
  Retrieval}} \bibinfo{volume}{8}, \bibinfo{number}{1} (\bibinfo{year}{2014}),
  \bibinfo{pages}{1--125}.
\newblock


\bibitem[\protect\citeauthoryear{G{\"u}tl, Pivec, Trummer, Garcia-Barrios,
  M{\"o}dritscher, Pripfl, and Umgeher}{G{\"u}tl et~al\mbox{.}}{2005}]%
        {gutl2005adele}
\bibfield{author}{\bibinfo{person}{Christian G{\"u}tl}, \bibinfo{person}{Maja
  Pivec}, \bibinfo{person}{Christian Trummer}, \bibinfo{person}{Victor~Manuel
  Garcia-Barrios}, \bibinfo{person}{Felix M{\"o}dritscher},
  \bibinfo{person}{Juergen Pripfl}, {and} \bibinfo{person}{Martin Umgeher}.}
  \bibinfo{year}{2005}\natexlab{}.
\newblock \showarticletitle{Adele (adaptive e-learning with eye-tracking):
  Theoretical background, system architecture and application scenarios}.
\newblock \bibinfo{journal}{\emph{European Journal of Open, Distance and
  E-Learning}} \bibinfo{volume}{8}, \bibinfo{number}{2} (\bibinfo{year}{2005}).
\newblock


\bibitem[\protect\citeauthoryear{Hardcastle, Blake, and Hagger}{Hardcastle
  et~al\mbox{.}}{2012}]%
        {hardcastle2012effectiveness}
\bibfield{author}{\bibinfo{person}{Sarah Hardcastle}, \bibinfo{person}{Nicola
  Blake}, {and} \bibinfo{person}{Martin~S Hagger}.}
  \bibinfo{year}{2012}\natexlab{}.
\newblock \showarticletitle{The effectiveness of a motivational interviewing
  primary-care based intervention on physical activity and predictors of change
  in a disadvantaged community}.
\newblock \bibinfo{journal}{\emph{Journal of behavioral medicine}}
  \bibinfo{volume}{35}, \bibinfo{number}{3} (\bibinfo{year}{2012}),
  \bibinfo{pages}{318--333}.
\newblock


\bibitem[\protect\citeauthoryear{Harrison}{Harrison}{2000}]%
        {harrison2000ebooks}
\bibfield{author}{\bibinfo{person}{B.L. Harrison}.}
  \bibinfo{year}{2000}\natexlab{}.
\newblock \showarticletitle{E-books and the future of reading}.
\newblock \bibinfo{journal}{\emph{IEEE Comput. Grap. Appl.}}
  \bibinfo{volume}{20}, \bibinfo{number}{3} (\bibinfo{year}{2000}),
  \bibinfo{pages}{32--39}.
\newblock


\bibitem[\protect\citeauthoryear{Henelius and Torniainen}{Henelius and
  Torniainen}{2018}]%
        {henelius2018}
\bibfield{author}{\bibinfo{person}{Andreas Henelius} {and}
  \bibinfo{person}{Jari Torniainen}.} \bibinfo{year}{2018}\natexlab{}.
\newblock \showarticletitle{{MIDAS}: Open-source framework for distributed
  online analysis of data streams}.
\newblock \bibinfo{journal}{\emph{SoftwareX}}  \bibinfo{volume}{7}
  (\bibinfo{year}{2018}), \bibinfo{pages}{156--161}.
\newblock


\bibitem[\protect\citeauthoryear{Hollingworth, Williams, and
  Henderson}{Hollingworth et~al\mbox{.}}{2001}]%
        {hollingworth2001see}
\bibfield{author}{\bibinfo{person}{Andrew Hollingworth},
  \bibinfo{person}{Carrick~C Williams}, {and} \bibinfo{person}{John~M
  Henderson}.} \bibinfo{year}{2001}\natexlab{}.
\newblock \showarticletitle{To see and remember: Visually specific information
  is retained in memory from previously attended objects in natural scenes}.
\newblock \bibinfo{journal}{\emph{Psychonomic Bulletin \& Review}}
  \bibinfo{volume}{8}, \bibinfo{number}{4} (\bibinfo{year}{2001}),
  \bibinfo{pages}{761--768}.
\newblock


\bibitem[\protect\citeauthoryear{Hoppe, Loetscher, Morey, and Bulling}{Hoppe
  et~al\mbox{.}}{2018}]%
        {hoppe18}
\bibfield{author}{\bibinfo{person}{Sabrina Hoppe}, \bibinfo{person}{Tobias
  Loetscher}, \bibinfo{person}{Stephanie~A. Morey}, {and}
  \bibinfo{person}{Andreas Bulling}.} \bibinfo{year}{2018}\natexlab{}.
\newblock \showarticletitle{Eye Movements During Everyday Behavior Predict
  Personality Traits}.
\newblock \bibinfo{journal}{\emph{Frontiers in Human Neuroscience}}
  \bibinfo{volume}{12} (\bibinfo{year}{2018}), \bibinfo{pages}{105}.
\newblock
\showISSN{1662-5161}
\urldef\tempurl%
\url{https://doi.org/10.3389/fnhum.2018.00105}
\showDOI{\tempurl}


\bibitem[\protect\citeauthoryear{Jacob and Karn}{Jacob and Karn}{2003}]%
        {jacob2003eye}
\bibfield{author}{\bibinfo{person}{Robert~JK Jacob} {and}
  \bibinfo{person}{Keith~S Karn}.} \bibinfo{year}{2003}\natexlab{}.
\newblock \showarticletitle{Eye tracking in human-computer interaction and
  usability research: Ready to deliver the promises}.
\newblock In \bibinfo{booktitle}{\emph{The mind's eye}}.
  \bibinfo{publisher}{Elsevier}, \bibinfo{pages}{573--605}.
\newblock


\bibitem[\protect\citeauthoryear{Jeni, Cohn, and De~La~Torre}{Jeni
  et~al\mbox{.}}{2013}]%
        {jeni2013facing}
\bibfield{author}{\bibinfo{person}{Laszlo~A. Jeni}, \bibinfo{person}{Jeffrey~F.
  Cohn}, {and} \bibinfo{person}{Fernando De~La~Torre}.}
  \bibinfo{year}{{2013}}\natexlab{}.
\newblock \showarticletitle{{Facing Imbalanced Data Recommendations for the Use
  of Performance Metrics}}. In \bibinfo{booktitle}{\emph{{2013 HUMAINE
  ASSOCIATION CONFERENCE ON AFFECTIVE COMPUTING AND INTELLIGENT INTERACTION
  (ACII)}}} \emph{(\bibinfo{series}{{International Conference on Affective
  Computing and Intelligent Interaction}})}. {Humaine Assoc; IEEE Comp Soc;
  Comp Vis; Multimedia Lab; Univ Geneva, Swiss Ctr Affect Sci; GFK Verein;
  Technicolor; Telono; Brain Prod; Inst Telecom; Telecom ParisTech; Swiss Natl
  Sci Fdn; Soc Academique Geneve; Amer Assoc Artificial Itelligence},
  \bibinfo{pages}{{245--251}}.
\newblock
\showISBNx{{978-0-7695-5048-0}}
\showISSN{{2156-8103}}
\urldef\tempurl%
\url{https://doi.org/{10.1109/ACII.2013.47}}
\showDOI{\tempurl}
\newblock
\shownote{{5th Biannual Conference of the Humaine-Association on Affective
  Computing and Intelligent Interaction (ACII), Geneva, SWITZERLAND, SEP 02-05,
  2013}.}


\bibitem[\protect\citeauthoryear{Koch, John, Worner, Muller, and Ertl}{Koch
  et~al\mbox{.}}{2014}]%
        {Koch2014}
\bibfield{author}{\bibinfo{person}{Steffen Koch}, \bibinfo{person}{Markus
  John}, \bibinfo{person}{Michael Worner}, \bibinfo{person}{Andreas Muller},
  {and} \bibinfo{person}{Thomas Ertl}.} \bibinfo{year}{2014}\natexlab{}.
\newblock \showarticletitle{VarifocalReader --- In-Depth Visual Analysis of
  Large Text Documents}.
\newblock \bibinfo{journal}{\emph{IEEE Trans. Visual. Comput. Graphics}}
  \bibinfo{volume}{20}, \bibinfo{number}{12} (\bibinfo{year}{2014}),
  \bibinfo{pages}{1723--1732}.
\newblock


\bibitem[\protect\citeauthoryear{Kunze, Ishimaru, Utsumi, and Kise}{Kunze
  et~al\mbox{.}}{2013a}]%
        {kunze2013my}
\bibfield{author}{\bibinfo{person}{Kai Kunze}, \bibinfo{person}{Shoya
  Ishimaru}, \bibinfo{person}{Yuzuko Utsumi}, {and} \bibinfo{person}{Koichi
  Kise}.} \bibinfo{year}{2013}\natexlab{a}.
\newblock \showarticletitle{My reading life}. In
  \bibinfo{booktitle}{\emph{Proceedings of the 2013 ACM conference on Pervasive
  and ubiquitous computing adjunct publication - UbiComp `13 Adjunct}}.
  \bibinfo{publisher}{ACM Press}, \bibinfo{address}{New York, New York, USA}.
\newblock


\bibitem[\protect\citeauthoryear{Kunze, Kawaichi, Yoshimura, and Kise}{Kunze
  et~al\mbox{.}}{2013b}]%
        {kunze2013towards}
\bibfield{author}{\bibinfo{person}{Kai Kunze}, \bibinfo{person}{Hitoshi
  Kawaichi}, \bibinfo{person}{Kazuyo Yoshimura}, {and} \bibinfo{person}{Koichi
  Kise}.} \bibinfo{year}{2013}\natexlab{b}.
\newblock \showarticletitle{Towards inferring language expertise using eye
  tracking}. In \bibinfo{booktitle}{\emph{CHI `13 Extended Abstracts on Human
  Factors in Computing Systems on - CHI EA `13}}. \bibinfo{publisher}{ACM
  Press}, \bibinfo{address}{New York, New York, USA}.
\newblock


\bibitem[\protect\citeauthoryear{Lall{\'e}, Conati, and Carenini}{Lall{\'e}
  et~al\mbox{.}}{2016}]%
        {lalle2016predicting}
\bibfield{author}{\bibinfo{person}{S{\'e}bastien Lall{\'e}},
  \bibinfo{person}{Cristina Conati}, {and} \bibinfo{person}{Giuseppe
  Carenini}.} \bibinfo{year}{2016}\natexlab{}.
\newblock \showarticletitle{Predicting Confusion in Information Visualization
  from Eye Tracking and Interaction Data.}. In
  \bibinfo{booktitle}{\emph{International Joint Conference on Artificial
  Intelligence}}. \bibinfo{pages}{2529--2535}.
\newblock


\bibitem[\protect\citeauthoryear{Li, Babcock, and Parkhurst}{Li
  et~al\mbox{.}}{2006}]%
        {li2006openeyes}
\bibfield{author}{\bibinfo{person}{Dongheng Li}, \bibinfo{person}{Jason
  Babcock}, {and} \bibinfo{person}{Derrick~J Parkhurst}.}
  \bibinfo{year}{2006}\natexlab{}.
\newblock \showarticletitle{openEyes: a low-cost head-mounted eye-tracking
  solution}. In \bibinfo{booktitle}{\emph{Proceedings of the 2006 symposium on
  Eye tracking research \& applications}}. \bibinfo{publisher}{ACM},
  \bibinfo{pages}{95--100}.
\newblock


\bibitem[\protect\citeauthoryear{Liu}{Liu}{2005}]%
        {Liu2005}
\bibfield{author}{\bibinfo{person}{Ziming Liu}.}
  \bibinfo{year}{2005}\natexlab{}.
\newblock \showarticletitle{Reading behavior in the digital environment:
  Changes in reading behavior over the past ten years}.
\newblock \bibinfo{journal}{\emph{Journal of documentation}}
  \bibinfo{volume}{61}, \bibinfo{number}{6} (\bibinfo{year}{2005}),
  \bibinfo{pages}{700--712}.
\newblock


\bibitem[\protect\citeauthoryear{Mangen, Walgermo, and Br{\o}nnick}{Mangen
  et~al\mbox{.}}{2013}]%
        {mangen2013}
\bibfield{author}{\bibinfo{person}{Anne Mangen}, \bibinfo{person}{Bente~R.
  Walgermo}, {and} \bibinfo{person}{Kolbj{\o}rn Br{\o}nnick}.}
  \bibinfo{year}{2013}\natexlab{}.
\newblock \showarticletitle{Reading linear texts on paper versus computer
  screen: Effects on reading comprehension}.
\newblock \bibinfo{journal}{\emph{International Journal of Educational
  Research}}  \bibinfo{volume}{58} (\bibinfo{year}{2013}),
  \bibinfo{pages}{61--68}.
\newblock


\bibitem[\protect\citeauthoryear{O'Hara, Smith, Newman, and Sellen}{O'Hara
  et~al\mbox{.}}{1998}]%
        {o1998student}
\bibfield{author}{\bibinfo{person}{Kenton O'Hara}, \bibinfo{person}{Fiona
  Smith}, \bibinfo{person}{William Newman}, {and} \bibinfo{person}{Abigail
  Sellen}.} \bibinfo{year}{1998}\natexlab{}.
\newblock \showarticletitle{Student readers' use of library documents:
  implications for library technologies Proceedings of the {SIGCHI} conference
  on Human factors in computing systems}. \bibinfo{publisher}{ACM
  Press/Addison-Wesley Publishing Co.}, \bibinfo{pages}{233--240}.
\newblock


\bibitem[\protect\citeauthoryear{Ojanp{\"a}{\"a}, N{\"a}s{\"a}nen, and
  Kojo}{Ojanp{\"a}{\"a} et~al\mbox{.}}{2002}]%
        {ojanpaa2002eye}
\bibfield{author}{\bibinfo{person}{Helena Ojanp{\"a}{\"a}},
  \bibinfo{person}{Risto N{\"a}s{\"a}nen}, {and} \bibinfo{person}{Ilpo Kojo}.}
  \bibinfo{year}{2002}\natexlab{}.
\newblock \showarticletitle{Eye movements in the visual search of word lists}.
\newblock \bibinfo{journal}{\emph{Vision Research}} \bibinfo{volume}{42},
  \bibinfo{number}{12} (\bibinfo{year}{2002}), \bibinfo{pages}{1499--1512}.
\newblock


\bibitem[\protect\citeauthoryear{Primack, Shensa, Kim, Carroll, Hoban, Leino,
  Eissenberg, Dachille, and Fine}{Primack et~al\mbox{.}}{2012}]%
        {primack2012waterpipe}
\bibfield{author}{\bibinfo{person}{Brian~A Primack}, \bibinfo{person}{Ariel
  Shensa}, \bibinfo{person}{Kevin~H Kim}, \bibinfo{person}{Mary~V Carroll},
  \bibinfo{person}{Mary~T Hoban}, \bibinfo{person}{E~Victor Leino},
  \bibinfo{person}{Thomas Eissenberg}, \bibinfo{person}{Kathleen~H Dachille},
  {and} \bibinfo{person}{Michael~J Fine}.} \bibinfo{year}{2012}\natexlab{}.
\newblock \showarticletitle{Waterpipe smoking among US university students}.
\newblock \bibinfo{journal}{\emph{Nicotine \& Tobacco Research}}
  \bibinfo{volume}{15}, \bibinfo{number}{1} (\bibinfo{year}{2012}),
  \bibinfo{pages}{29--35}.
\newblock


\bibitem[\protect\citeauthoryear{R.-Tavakoli, Atyabi, Rantanen, Laukka,
  Nefti-Meziani, and Heikkilä}{R.-Tavakoli et~al\mbox{.}}{2015}]%
        {Tavakoli2015}
\bibfield{author}{\bibinfo{person}{Hamed R.-Tavakoli}, \bibinfo{person}{Adham
  Atyabi}, \bibinfo{person}{Antti Rantanen}, \bibinfo{person}{Seppo~J. Laukka},
  \bibinfo{person}{Samia Nefti-Meziani}, {and} \bibinfo{person}{Janne
  Heikkilä}.} \bibinfo{year}{2015}\natexlab{}.
\newblock \showarticletitle{Predicting the Valence of a Scene from Observers’
  Eye Movements}.
\newblock \bibinfo{journal}{\emph{PLOS ONE}} \bibinfo{volume}{10},
  \bibinfo{number}{9} (\bibinfo{date}{09} \bibinfo{year}{2015}),
  \bibinfo{pages}{1--19}.
\newblock
\urldef\tempurl%
\url{https://doi.org/10.1371/journal.pone.0138198}
\showDOI{\tempurl}


\bibitem[\protect\citeauthoryear{R.-Tavakoli, Poostchi, Peltonen, Laaksonen,
  and Kaski}{R.-Tavakoli et~al\mbox{.}}{2016}]%
        {Tavakoli16}
\bibfield{author}{\bibinfo{person}{Hamed R.-Tavakoli}, \bibinfo{person}{Hanieh
  Poostchi}, \bibinfo{person}{Jaakko Peltonen}, \bibinfo{person}{Jorma
  Laaksonen}, {and} \bibinfo{person}{Samuel Kaski}.}
  \bibinfo{year}{2016}\natexlab{}.
\newblock \showarticletitle{Preliminary Studies on Personalized Preference
  Prediction from Gaze in Comparing Visualizations}. In
  \bibinfo{booktitle}{\emph{Advances in Visual Computing}},
  \bibfield{editor}{\bibinfo{person}{George Bebis}, \bibinfo{person}{Richard
  Boyle}, \bibinfo{person}{Bahram Parvin}, \bibinfo{person}{Darko Koracin},
  \bibinfo{person}{Fatih Porikli}, \bibinfo{person}{Sandra Skaff},
  \bibinfo{person}{Alireza Entezari}, \bibinfo{person}{Jianyuan Min},
  \bibinfo{person}{Daisuke Iwai}, \bibinfo{person}{Amela Sadagic},
  \bibinfo{person}{Carlos Scheidegger}, {and} \bibinfo{person}{Tobias
  Isenberg}} (Eds.). \bibinfo{publisher}{Springer International Publishing},
  \bibinfo{address}{Cham}, \bibinfo{pages}{576--585}.
\newblock
\showISBNx{978-3-319-50832-0}


\bibitem[\protect\citeauthoryear{Rayner, Chace, Slattery, and Ashby}{Rayner
  et~al\mbox{.}}{2006a}]%
        {rayner2006comprehension}
\bibfield{author}{\bibinfo{person}{Keith Rayner}, \bibinfo{person}{Kathryn~H
  Chace}, \bibinfo{person}{Timothy~J Slattery}, {and} \bibinfo{person}{Jane
  Ashby}.} \bibinfo{year}{2006}\natexlab{a}.
\newblock \showarticletitle{Eye Movements as Reflections of Comprehension
  Processes in Reading}.
\newblock \bibinfo{journal}{\emph{Scientific Studies of Reading}}
  \bibinfo{volume}{10}, \bibinfo{number}{3} (\bibinfo{year}{2006}),
  \bibinfo{pages}{241--255}.
\newblock


\bibitem[\protect\citeauthoryear{Rayner, Chace, Slattery, and Ashby}{Rayner
  et~al\mbox{.}}{2006b}]%
        {rayner2006eye}
\bibfield{author}{\bibinfo{person}{Keith Rayner}, \bibinfo{person}{Kathryn~H
  Chace}, \bibinfo{person}{Timothy~J Slattery}, {and} \bibinfo{person}{Jane
  Ashby}.} \bibinfo{year}{2006}\natexlab{b}.
\newblock \showarticletitle{Eye movements as reflections of comprehension
  processes in reading}.
\newblock \bibinfo{journal}{\emph{Scientific Studies of Reading}}
  \bibinfo{volume}{10}, \bibinfo{number}{3} (\bibinfo{year}{2006}),
  \bibinfo{pages}{241--255}.
\newblock


\bibitem[\protect\citeauthoryear{Richardson, Dale, and Spivey}{Richardson
  et~al\mbox{.}}{2007}]%
        {gonzalezmarquez2007eye}
\bibfield{author}{\bibinfo{person}{Daniel~C. Richardson}, \bibinfo{person}{Rick
  Dale}, {and} \bibinfo{person}{Michael~J. Spivey}.}
  \bibinfo{year}{2007}\natexlab{}.
\newblock \bibinfo{booktitle}{\emph{Eye Movements in Language and Cognition}}.
\newblock \bibinfo{publisher}{John Benjamins Publishing Company}.
\newblock


\bibitem[\protect\citeauthoryear{Rose, Geers, Rasinski, and Fowler}{Rose
  et~al\mbox{.}}{2012}]%
        {rose2012choice}
\bibfield{author}{\bibinfo{person}{Jason~P Rose}, \bibinfo{person}{Andrew~L
  Geers}, \bibinfo{person}{Heather~M Rasinski}, {and}
  \bibinfo{person}{Stephanie~L Fowler}.} \bibinfo{year}{2012}\natexlab{}.
\newblock \showarticletitle{Choice and placebo expectation effects in the
  context of pain analgesia}.
\newblock \bibinfo{journal}{\emph{Journal of behavioral medicine}}
  \bibinfo{volume}{35}, \bibinfo{number}{4} (\bibinfo{year}{2012}),
  \bibinfo{pages}{462--470}.
\newblock


\bibitem[\protect\citeauthoryear{Santini, Fuhl, K{\"u}bler, and
  Kasneci}{Santini et~al\mbox{.}}{2016}]%
        {santini2016eyerec}
\bibfield{author}{\bibinfo{person}{Thiago Santini}, \bibinfo{person}{Wolfgang
  Fuhl}, \bibinfo{person}{Thomas K{\"u}bler}, {and} \bibinfo{person}{Enkelejda
  Kasneci}.} \bibinfo{year}{2016}\natexlab{}.
\newblock \showarticletitle{Eyerec: An open-source data acquisition software
  for head-mounted eye-tracking}. In \bibinfo{booktitle}{\emph{International
  Conference on Vision Theory and Applications (VISAPP)}}.
\newblock


\bibitem[\protect\citeauthoryear{Schilit, Golovchinsky, and Price}{Schilit
  et~al\mbox{.}}{1998}]%
        {Schilit1998}
\bibfield{author}{\bibinfo{person}{Bill~N Schilit}, \bibinfo{person}{Gene
  Golovchinsky}, {and} \bibinfo{person}{Morgan~N Price}.}
  \bibinfo{year}{1998}\natexlab{}.
\newblock \showarticletitle{Beyond paper: supporting active reading with free
  form digital ink annotations}. In \bibinfo{booktitle}{\emph{Proceedings of
  the SIGCHI conference on Human factors in computing systems}}.
  \bibinfo{publisher}{ACM Press/Addison-Wesley Publishing Co.},
  \bibinfo{pages}{249--256}.
\newblock


\bibitem[\protect\citeauthoryear{Schneider and Pea}{Schneider and Pea}{2013}]%
        {schneider2013}
\bibfield{author}{\bibinfo{person}{Bertrand Schneider} {and}
  \bibinfo{person}{Roy Pea}.} \bibinfo{year}{2013}\natexlab{}.
\newblock \showarticletitle{Real-time mutual gaze perception enhances
  collaborative learning and collaboration quality}.
\newblock \bibinfo{journal}{\emph{Intern. J. Comput.-Support. Collab. Learn.}}
  \bibinfo{volume}{8}, \bibinfo{number}{4} (\bibinfo{year}{2013}),
  \bibinfo{pages}{375--397}.
\newblock


\bibitem[\protect\citeauthoryear{Shaffer, Wise, Walters, M{\"u}ller, Falcone,
  and Sharif}{Shaffer et~al\mbox{.}}{2015}]%
        {shaffer2015itrace}
\bibfield{author}{\bibinfo{person}{Timothy~R. Shaffer},
  \bibinfo{person}{Jenna~L. Wise}, \bibinfo{person}{Braden~M. Walters},
  \bibinfo{person}{Sebastian~C. M{\"u}ller}, \bibinfo{person}{Michael Falcone},
  {and} \bibinfo{person}{Bonita Sharif}.} \bibinfo{year}{2015}\natexlab{}.
\newblock \showarticletitle{iTrace: enabling eye tracking on software artifacts
  within the {IDE} to support software engineering tasks}. In
  \bibinfo{booktitle}{\emph{Proceedings of the 2015 10th Joint Meeting on
  Foundations of Software Engineering - ESEC/FSE 2015}}.
  \bibinfo{publisher}{ACM Press}, \bibinfo{address}{New York, New York, USA}.
\newblock


\bibitem[\protect\citeauthoryear{Sharma, Jermann, and Dillenbourg}{Sharma
  et~al\mbox{.}}{2015}]%
        {sharma2015displaying}
\bibfield{author}{\bibinfo{person}{Kshitij Sharma}, \bibinfo{person}{Patrick
  Jermann}, {and} \bibinfo{person}{Pierre Dillenbourg}.}
  \bibinfo{year}{2015}\natexlab{}.
\newblock \showarticletitle{Displaying Teacher's Gaze in a MOOC: Effects on
  Students' Video Navigation Patterns}.
\newblock In \bibinfo{booktitle}{\emph{Design for Teaching and Learning in a
  Networked World}}. \bibinfo{publisher}{Springer}, \bibinfo{pages}{325--338}.
\newblock


\bibitem[\protect\citeauthoryear{Sj{\"o}berg, Chen, Flor{\'e}en, Koskela,
  Kuikkaniemi, Lehtiniemi, and Peltonen}{Sj{\"o}berg et~al\mbox{.}}{2016}]%
        {symbiotic2016}
\bibfield{author}{\bibinfo{person}{Mats Sj{\"o}berg}, \bibinfo{person}{Hung-Han
  Chen}, \bibinfo{person}{Patrik Flor{\'e}en}, \bibinfo{person}{Markus
  Koskela}, \bibinfo{person}{Kai Kuikkaniemi}, \bibinfo{person}{Tuukka
  Lehtiniemi}, {and} \bibinfo{person}{Jaakko Peltonen}.}
  \bibinfo{year}{2016}\natexlab{}.
\newblock \showarticletitle{Digital Me: Controlling and Making Sense of My
  Digital Footprint}. In \bibinfo{booktitle}{\emph{Proceedings of the 5th
  International Workshop on Symbiotic Interaction}}.
\newblock


\bibitem[\protect\citeauthoryear{Sogo}{Sogo}{2013a}]%
        {sogo2013agazeparser}
\bibfield{author}{\bibinfo{person}{H Sogo}.} \bibinfo{year}{2013}\natexlab{a}.
\newblock \showarticletitle{GazeParser: an open-source and multiplatform
  library for low-cost eye tracking and analysis.}
\newblock \bibinfo{journal}{\emph{Behavior Research Methods}}
  \bibinfo{volume}{45}, \bibinfo{number}{3} (\bibinfo{year}{2013}),
  \bibinfo{pages}{684--695}.
\newblock


\bibitem[\protect\citeauthoryear{Sogo}{Sogo}{2013b}]%
        {sogo2013}
\bibfield{author}{\bibinfo{person}{H Sogo}.} \bibinfo{year}{2013}\natexlab{b}.
\newblock \showarticletitle{GazeParser: an open-source and multiplatform
  library for low-cost eye tracking and analysis.}
\newblock \bibinfo{journal}{\emph{Behav Res Methods}} \bibinfo{volume}{45},
  \bibinfo{number}{3} (\bibinfo{year}{2013}), \bibinfo{pages}{684--695}.
\newblock


\bibitem[\protect\citeauthoryear{Stein and Brennan}{Stein and Brennan}{2004}]%
        {stein2004another}
\bibfield{author}{\bibinfo{person}{Randy Stein} {and} \bibinfo{person}{Susan~E
  Brennan}.} \bibinfo{year}{2004}\natexlab{}.
\newblock \showarticletitle{Another person's eye gaze as a cue in solving
  programming problems}. In \bibinfo{booktitle}{\emph{Proceedings of the 6th
  international conference on Multimodal interfaces}}.
  \bibinfo{publisher}{ACM}, \bibinfo{pages}{9--15}.
\newblock


\bibitem[\protect\citeauthoryear{Stewart, Turnbull, Mikocka-Walus, Harley, and
  Andrews}{Stewart et~al\mbox{.}}{2013}]%
        {stewart2013acceptability}
\bibfield{author}{\bibinfo{person}{Benjamin~JR Stewart},
  \bibinfo{person}{Deborah Turnbull}, \bibinfo{person}{Antonina~A
  Mikocka-Walus}, \bibinfo{person}{Hugh~AJ Harley}, {and}
  \bibinfo{person}{Jane~M Andrews}.} \bibinfo{year}{2013}\natexlab{}.
\newblock \showarticletitle{Acceptability of psychotherapy, pharmacotherapy,
  and self-directed therapies in Australians living with chronic hepatitis C}.
\newblock \bibinfo{journal}{\emph{Journal of clinical psychology in medical
  settings}} \bibinfo{volume}{20}, \bibinfo{number}{4} (\bibinfo{year}{2013}),
  \bibinfo{pages}{427--439}.
\newblock


\bibitem[\protect\citeauthoryear{Taipale}{Taipale}{2014}]%
        {taipale2014}
\bibfield{author}{\bibinfo{person}{Sakari Taipale}.}
  \bibinfo{year}{2014}\natexlab{}.
\newblock \showarticletitle{The affordances of reading/writing on paper and
  digitally in Finland}.
\newblock \bibinfo{journal}{\emph{Telematics and Informatics}}
  \bibinfo{volume}{31}, \bibinfo{number}{4} (\bibinfo{year}{2014}),
  \bibinfo{pages}{532--542}.
\newblock


\bibitem[\protect\citeauthoryear{Traxler, Morris, and Seely}{Traxler
  et~al\mbox{.}}{2002}]%
        {traxler2002clauses}
\bibfield{author}{\bibinfo{person}{Matthew~J Traxler}, \bibinfo{person}{Robin~K
  Morris}, {and} \bibinfo{person}{Rachel~E Seely}.}
  \bibinfo{year}{2002}\natexlab{}.
\newblock \showarticletitle{Processing Subject and Object Relative Clauses:
  Evidence from Eye Movements}.
\newblock \bibinfo{journal}{\emph{Journal of Memory and Language}}
  \bibinfo{volume}{47}, \bibinfo{number}{1} (\bibinfo{year}{2002}),
  \bibinfo{pages}{69--90}.
\newblock


\bibitem[\protect\citeauthoryear{van Gog and Scheiter}{van Gog and
  Scheiter}{2010}]%
        {vangog2010}
\bibfield{author}{\bibinfo{person}{Tamara van Gog} {and}
  \bibinfo{person}{Katharina Scheiter}.} \bibinfo{year}{2010}\natexlab{}.
\newblock \showarticletitle{Eye tracking as a tool to study and enhance
  multimedia learning}.
\newblock \bibinfo{journal}{\emph{Learning and Instruction}}
  \bibinfo{volume}{20}, \bibinfo{number}{2} (\bibinfo{year}{2010}),
  \bibinfo{pages}{95--99}.
\newblock


\bibitem[\protect\citeauthoryear{Vertegaal}{Vertegaal}{1999}]%
        {vertegaal1999gaze}
\bibfield{author}{\bibinfo{person}{Roel Vertegaal}.}
  \bibinfo{year}{1999}\natexlab{}.
\newblock \showarticletitle{The {GAZE} groupware system: mediating joint
  attention in multiparty communication and collaboration}. In
  \bibinfo{booktitle}{\emph{Proceedings of the SIGCHI conference on Human
  Factors in Computing Systems}}. \bibinfo{publisher}{ACM},
  \bibinfo{pages}{294--301}.
\newblock


\bibitem[\protect\citeauthoryear{Vossk{\"u}hler, Nordmeier, Kuchinke, and
  Jacobs}{Vossk{\"u}hler et~al\mbox{.}}{2008}]%
        {vosskuhler2008}
\bibfield{author}{\bibinfo{person}{A Vossk{\"u}hler}, \bibinfo{person}{V
  Nordmeier}, \bibinfo{person}{L Kuchinke}, {and} \bibinfo{person}{AM Jacobs}.}
  \bibinfo{year}{2008}\natexlab{}.
\newblock \showarticletitle{{OGAMA} (Open Gaze and Mouse Analyzer): open-source
  software designed to analyze eye and mouse movements in slideshow study
  designs.}
\newblock \bibinfo{journal}{\emph{Behav Res Methods}} \bibinfo{volume}{40},
  \bibinfo{number}{4} (\bibinfo{year}{2008}), \bibinfo{pages}{1150--1162}.
\newblock


\bibitem[\protect\citeauthoryear{Wilson and Russell}{Wilson and
  Russell}{2003}]%
        {wilson2003real}
\bibfield{author}{\bibinfo{person}{Glenn~F Wilson} {and}
  \bibinfo{person}{Christopher~A Russell}.} \bibinfo{year}{2003}\natexlab{}.
\newblock \showarticletitle{Real-Time Assessment of Mental Workload Using
  Psychophysiological Measures and Artificial Neural Networks}.
\newblock \bibinfo{journal}{\emph{Human Factors}} \bibinfo{volume}{45},
  \bibinfo{number}{4} (\bibinfo{year}{2003}), \bibinfo{pages}{635--643}.
\newblock


\bibitem[\protect\citeauthoryear{Wood and Bulling}{Wood and Bulling}{2014}]%
        {wood2014eyetab}
\bibfield{author}{\bibinfo{person}{Erroll Wood} {and} \bibinfo{person}{Andreas
  Bulling}.} \bibinfo{year}{2014}\natexlab{}.
\newblock \showarticletitle{Eyetab: Model-based gaze estimation on unmodified
  tablet computers}. In \bibinfo{booktitle}{\emph{Proceedings of the Symposium
  on Eye Tracking Research and Applications}}. \bibinfo{publisher}{ACM},
  \bibinfo{pages}{207--210}.
\newblock


\bibitem[\protect\citeauthoryear{Zhang and Meur}{Zhang and Meur}{2018}]%
        {Oli18}
\bibfield{author}{\bibinfo{person}{A.~T. Zhang} {and} \bibinfo{person}{B.~O.~Le
  Meur}.} \bibinfo{year}{2018}\natexlab{}.
\newblock \showarticletitle{How Old Do You Look? Inferring Your Age from Your
  Gaze}. In \bibinfo{booktitle}{\emph{2018 25th IEEE International Conference
  on Image Processing (ICIP)}}. \bibinfo{pages}{2660--2664}.
\newblock
\showISSN{2381-8549}
\urldef\tempurl%
\url{https://doi.org/10.1109/ICIP.2018.8451219}
\showDOI{\tempurl}


\end{thebibliography}

\end{document}